\documentclass[a4paper,11pt]{article}
\pdfoutput=1

\usepackage{jheppub}
\usepackage[T1]{fontenc} 
\usepackage{bm}
\usepackage{mathrsfs}
\usepackage{bbold}
\usepackage{amsbsy}
\usepackage{braket}
\usepackage{amssymb}
\usepackage{diagbox}
\usepackage{multirow}
\usepackage{array}
\usepackage{graphicx}
\usepackage[center]{subfigure}
\usepackage{subfigure}
\usepackage{amsmath}
\usepackage{amsfonts}
\usepackage{xcolor}
\usepackage{makecell}

\newcommand{\mpref}[1]{Figure.\ref{#1}}
\newcommand{\be}{\begin{equation}}
\newcommand{\ee}{\end{equation}}
\newcommand{\bea}{\begin{eqnarray}}
\newcommand{\eea}{\end{eqnarray}}

\title{\boldmath Replica Wormholes, Modular Entropy, and Capacity of Entanglement in JT Gravity}

\author[a]{Ming-Hui Yu,}
\author[a]{Shu-Yi Lin,}
\author[a,1]{Xian-Hui Ge,\note{Corresponding author.}}

\affiliation[a]{Department of Physics, College of Sciences, Shanghai University,\\99 Shangda Road, 200444 Shanghai, China}
\emailAdd{yuminghui@shu.edu.cn}
\emailAdd{linshuyi@shu.edu.cn}
\emailAdd{gexh@shu.edu.cn}

\abstract{By employing the replica trick we study the impact of the replica parameter $n$ on the modular entropy and the capacity of entanglement in the End of the World (EoW) model and the island model, respectively.  For the EoW model, we present $n$-dependent evolution curves of the modular entropy and the capacity of entanglement under both microcanonical and canonical ensembles. In particular, in the canonical ensemble, all quantities decrease as $n$ increases at late times. For the island model, we develop the replica geometry for finite $n$ and re-evaluate the modular entropy and the capacity of entanglement in a two-sided eternal Jackiw-Teitelboim black hole coupled with a thermal bath. In the case of a single island configuration, the modular entropy and capacity of entanglement closely resemble the thermal entropy and the heat capacity, respectively, yielding results analogous to those obtained in the canonical ensemble for the EoW model. The analysis of the results from these two models strongly indicates that in geometries with a greater number of $n$ copies, more connected geometries effectively purify thermal Hawking radiation. In addition, we compare these findings with statistical mechanics and provide an interpretation for the replica parameter $n$. Finally, we generalize the island formula to accommodate the finite $n$ case under this interpretation.}

\begin{document}
\maketitle
\flushbottom

\section{Introduction} \label{intro}
\quad Black holes play a highly particular role in modern physics. They are not only as existing components of the universe we live in, but also as an ideal testing subjects for the theory of quantum gravity, which attempts to unify general relativity and quantum mechanics. When quantum mechanics is inserted in the background of black holes, they demonstrate some non-trivial thermodynamic characteristics \cite{HR}. However, it is precisely the thermodynamic properties of black holes that present a long-standing challenge to theoretical physics, which known as the information loss paradox \cite{paradox}.

\par Recently, there has been a significant breakthrough on this issue \cite{bulk entropy,island rule,entanglement wedge,review}. The QES (quantum extremal surfaces) prescription \cite{QES} has been utilized in an evaporating black hole \cite{bulk entropy,entanglement wedge}, and the entanglement entropy of Hawking radiation in accordance with the Page curve \cite{PC1,PC2} is obtained. One can summarize the following island rule/formula based on that \cite{island rule,review},
\begin{equation}
S_R = \text{min} \bigg[ \text{ext} \bigg( \frac{\text{Area}(\partial I)}{4G_N} + S_{\text{bulk}} (I \cup R) \bigg) \bigg]. \label{island formula1}
\end{equation}
The expression in parentheses is called the generalized entropy of radiation, where the first term is the area term. It is an extension of the RT \cite{RT}/HRT \cite{HRT}/QRT \cite{QRT}formula. The contribution in the second term comes from the entropy of conformal field theory (CFT) in the bulk spacetime.  It should be noted that this entropy comprises two intervals. One is the region of radiation $R$ outside the black hole, and the other is the region of island $I$ \emph{inside} the black hole (its boundary is denoted as $\partial I$). This is the great contribution of the island formula. The island inside the black hole is connected in the gravitational region as a component of the entanglement region in the calculation of the entanglement entropy of Hawking radiation. A more precise description is the entanglement wedge reconstruction \cite{entanglement wedge}, namely, the \emph{internal} entanglement wedge of black holes is located in the entanglement wedge of the \emph{external} radiation. See an excellent review \cite{review}.

\par In addition, the island formula \eqref{island formula1} seems to be a direct application of the QES prescription \cite{QES}. However, a mathematically precise derivation of the path integral that yields the desired outcome, which is called the replica wormholes method \cite{replica1,replica2,replica3} and is written as follows
\begin{equation}
S_R=\lim\limits_{n \to 1} \frac{1}{1-n} \log \big[ \text{Tr} (\rho_R^n) \big]. \label{replica wormholes method}
\end{equation}
Here $\rho_R$ is the reduced density matrix of the state of radiation. As we shall see later, in the language of the path integral, Hawking ignored some \emph{possible paths}. Although these paths are trivial during the initial stage of the evaporation, they dominate in the final stage, which leads to a unitary Page curve.

\par At present, reproducing the Page curve in a black hole model involving Hawking radiation remains a challenging task. We have accomplished quite a lot thus far, such as some other two-dimensional (2D) black hole background \cite{eternalbh,jt1,2d1,2d2,2d3,2d4,2d5,2d6,2d7,2d8,2d9}; de Sitter spacetime \cite{ds1,ds2,ds3,ds4,ds5}; cosmology \cite{cosmology1,cosmology2,cosmology3,cosmology4,cosmology5,cosmology6}; gravitational bath \cite{grav bath1,grav bath2,grav bath3}; complexity \cite{complexity1,complexity2,complexity3,complexity4,complexity5,complexity6}; wedge holography \cite{wedge holo1,wedge holo2,wedge holo3,wedge holo4,wedge holo5,wedge holo6}; entanglement negativity \cite{negative1,negative2,negative3,negative4,negative5}; reflected entropy \cite{reflected entropy1,reflected entropy2}; baby universe \cite{baby universe1,baby universe2,baby universe3,baby universe4,baby universe5,baby universe6,baby universe7,baby universe8}; bra-ket wormholes \cite{braket1,braket2}; Sachedev-Ye-Kitaev model \cite{replica1,syk1,syk2,syk3,syk4,syk5}; moving mirror \cite{mirror1,mirror2}, and some other related work\cite{coe1,coe2,renyi wormholes,other1,other2,other3,other4,other5,other6,other7,other8,other9,other10,other11,other12,other13,other14,other15,other16,other17,other18,other19,other20,other21,other22,other23,other24,other25,other26,other27,other28,other29,other30,other31,other32,other33,other34,other35,other36,other37,other38,other39,other40,other41,other42,ge1,ge2}.

\par It should be noted that almost all current work involving the replica wormholes method only calculate the corresponding von Neumann entropy of radiation in the limit of $n \to 1$ \cite{replica1,replica2,replica3,coe1,coe2}. Due to the fact that the von Neumann entropy only offers an overall characteristic of the state, particularly the entanglement of the state. In order to obtain more information of states, we also need to adopt other measures, such as the R\'enyi entropy $S_{n}$ \cite{renyi entropy} , or more precisely, the modular entropy $S_{\text{mod}}$ \cite{modular entropy}
\begin{subequations}
\begin{align}
S_n &\equiv \frac{1}{1-n} \log \big[ \text{Tr} (\rho^n)  \big].  \label{renyi entropy1} \\
S_{\text{mod}} &\equiv n^2 \partial_n \bigg( \frac{n-1}{n} S_n\bigg)= S_n +n(n-1) \partial_n S_n. \label{modular entropy1}
\end{align}
\end{subequations}
At the same time, we can derive the capacity of entanglement (CoE) as a by-product, defined as the first derivative of the modular entropy with respect to the replica parameter $n$,
\begin{equation}
C_n =-\partial_n S_{\text{mod}} = n^2 \partial_n^2 [(1-n) S_n]. \label{entanglement capacity}
\end{equation}
Originally, the CoE was employed to describe topologically ordered states in condensed matter physics \cite{coe cmt}. It is a relevant quantity for measuring entanglement. Namely, it provides us with information about quantum entanglement that differs from the entanglement entropy. In particular, it appears in some recent studies of quantum gravity \cite{coe1,coe2,modular entropy,coe volume law,coe scalar,coe operator,coe gaussain}.

\par Therefore, the first motivation of this paper is to extend the previous work to the general case of the replica parameter $n$. Specifically, we will work in two heuristic models\footnote{Naively, there is no significant distinction between the EoW model and the island model when we only consider the contribution of the Euclidean path integral to the Page curve. The disconnected and connected geometries in the former correspond to the without-island and with-island configurations of the island model \cite{review}.}: the End of the World (EoW) model \cite{replica1} and the island model \cite{replica2,replica3}. Both can describe the black hole evaporation. We first calculate the time-dependent evolution of the modular entropy and the CoE under the microcanonical ensemble and the canonical ensemble, respectively through the gravitational path integral for the EoW model. Particularly, in the canonical ensemble, the numerical calculation shows that the modular entropy and the CoE decrease with the increase of $n$. Next, we develop the island model to the finite $n$ case and obtain the replica geometry. We finally obtain the analytical expressions for the modular entropy and the CoE. Surprisingly, in the configuration of single island, the modular entropy and the CoE also decreases as the parameter $n$ increases. This results are in agreement with the numerical results in the EoW model, which also indicates that the emergence of islands at late times is the key to the dominance of the connected geometry. By summarizing the findings from these two models, we conclude that for replica geometries with a larger $n$ copy number, an increase in the quantity of the connected geometry or the structure with islands will result in an increased purification of Hawking radiation.

\par The second motivation of the paper is to explore the physical significance of the replica parameter $n$ based on our results. It is widely recognized that by introducing the modular Hamiltonian, we can construct the corresponding physical quantity through analogy with statistical mechanics, such as the thermal entropy and the heat capacity. This analogy leads to the definition of new physical quantities, such as the modular entropy and the CoE. Refer to Table \ref{table1} for an analogy between statistical mechanics and the replica trick method. They can provide us with a better comprehension of the correlation between quantum entanglement and spacetime: The study of the modular entropy initially emerged in the study on the relationship between the AdS/CFT duality \cite {adscft} and the geometry of spacetime \cite{modular holo}. It is important to note that our calculation indicates that the parameter $n$ and the inverse temperature $\beta$ are consistently coupled and jointly manifest in the expressions for the modular entropy and the CoE. In addition, the modular entropy and the CoE at late times are analogy to the thermal entropy and the heat capacity, respectively. It is strongly suggested that the correspondence in the Table \ref{table1} does not merely represent a superficial analogy, but rather uncovers a profound and intrinsic connection between quantum information and statistical mechanics. Based on these results, we offer a reasonable explanation for the replica parameter $n$. Under this interpretation, the island formula \eqref{island formula1} can be generalized to the case of finite $n$ for JT gravity
\begin{equation}
S_R (n)  = \text{min}  \bigg[ \text{ext} \bigg ( \sum_{\partial I} (S_0 + \phi_n (\partial I)) +S_{\text{mod}} (R \cup I) \bigg) \bigg]. \label{island formula n}
\end{equation}
The original island formula \eqref{island formula1} is recovered in the limit when $n\to1$.

\begin{table}[htb]
\centering
 \linespread{2}\selectfont
\setlength{\tabcolsep}{0.5mm}
   \begin{tabular}{|c|c|}
    \hline
     \textbf{Statistical Mechanics} & \textbf{Replica Trick Method} \\ \hline
     Inverse Temperature $\beta$ & Replica Parameter $n$  \\ \hline
     Hamiltonian $\mathcal{H}$ & Modular Hamiltonian $\mathcal{H}_{\text{mod}}=-\log (\rho)$  \\ \hline
     Partition Function $Z(\beta) = \text{Tr} (e^{-\beta \mathcal{H}})$  & Replica Partition Function $ Z(n)= \text{Tr} (e^{-n \mathcal{H}_{\text{mod}}})$ \\ \hline
     Free Energy $F(\beta) =- \frac{1}{\beta} \log \big( Z(\beta) \big) $ & Replica Free Energy $F(n)=-\frac{1}{n}\log \big( Z(n) \big) $ \\ \hline
     \makecell{Thermal/Course-grained Entropy \\ $S(\beta)=\beta ^2 \frac{\partial F(\beta)}{\partial \beta}$} & \makecell{\textbf{Modular Entropy} \\ $S_{\text{mod}}(n)=n^2 \frac{\partial F(n)}{\partial n}$} \\ \hline
     Thermal Capacity $C(\beta)=-\beta \frac{\partial S(\beta)}{\partial \beta}$& Capacity of Entanglement $C_n =-n \frac{\partial S_{\text{mod}}(n)}{\partial n}$ \\ \hline
     \end{tabular}.
   \caption{The analogy between statistical mechanics and the replica trick method. It is essential to note that the analogous concept for the thermal entropy is the \emph{modular entropy}, rather than the R\'enyi entropy. We will elaborate on this in greater detail in the subsequent sections.}
    \label{table1}
\end{table}

\par The subsequent part of the paper is organized as follows. In section \ref{sec2}, we initially review the EoW model and explain the replica trick in the language of the path integral. Subsequently, we calculate the time-dependent evolution curves of the modular entropy and the CoE under two ensembles, respectively. In section \ref{sec3}, we shift our focus to the island model. Firstly, we calculate the analytical expressions of the modular entropy and the CoE through the gravitational path integral in the island model. In the specific calculation, we resort the high temperature limit to consider the $n$-dependence of the conformal welding problem. Then, we evaluate the modular entropy and the CoE in the eternal two-sided JT black hole with the single QES configuration. Eventually, we find that the results have a degree of consistency with the numerical results in section \ref{sec2}. Comparing the findings from these two models, we provide a reasonable interpretation of the replica parameter $n$. Based on this explanation, we can extend the original island formula to the case of finite $n$ for JT gravity. The discussion and the conclusion are presented in section \ref{sec4}. Three appendices \ref{appendixa}, \ref{appendixb}, and \ref{appendixc} provide additional details of the calculations.

\section{Replica Parameter Corrections to the EoW Model} \label{sec2}
\quad In this section, we calculate the time-dependent evolution of the modular entropy and the CoE in two kinds of ensembles in the EoW model. Since our final results are $n$-dependent, this extends to the previous work \cite{replica1} and brings several interesting innovations (\mpref{modular microcanonical}, \mpref{coe microcanonical}, \mpref{modular canonical} and \mpref{coe canonical}). We consider the EoW model as the initial simple toy model. It represents a 2D gravitational system describing an evaporating black hole in JT gravity with an EoW brane coupled to the auxiliary system \cite{replica1}. This model can provide us with an intuitive and insightful understanding for the island in the subsequent island model. Namely, the appearance of islands at late times is equivalent to the dominance of the replica wormholes saddle in the connected geometry at the late stage. Another advantage of this model is that in the limit of the number of large dimension $k$, the resolvent equation will reduce to a simple solvable form. Accordingly, the spectral function can be solved analytically in the microcanonical ensemble and numerically in the canonical ensemble.

\subsection{A Brief Review} \label{eow model}
\begin{figure}[htb]
\centering
\includegraphics[scale=0.2]{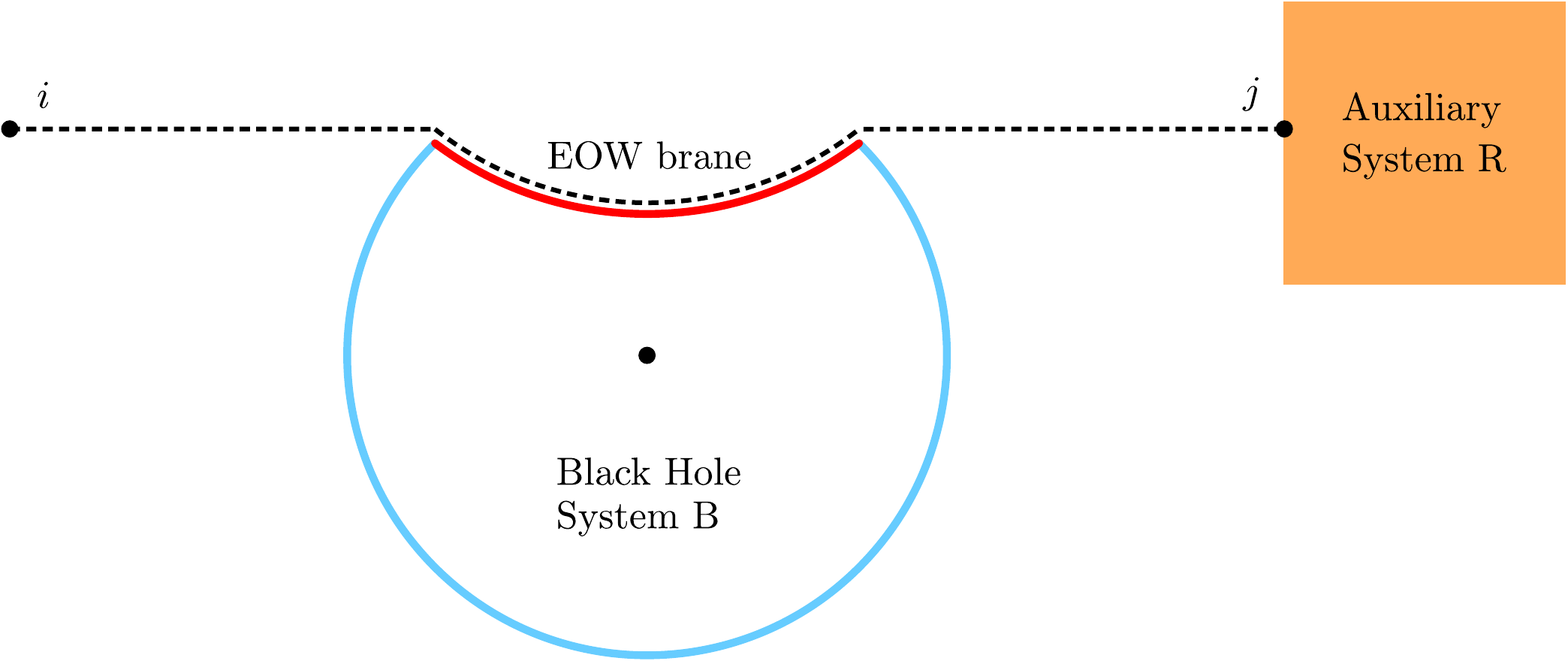}
\caption{\label{eow} The schematic diagram for the EoW model. We also present the geometry of  $\braket{\psi_i |\psi_j}$ by the dashed line. The hyperbolic disk (blue) represents an AdS$_2$, which takes the EoW brane (red) as the asymptotic boundary.}
\end{figure}

\quad In order to facilitate the subsequent discussion of the influence of the replica parameter $n$ and the utilization of notations in the gravitational path integral, we review the EoW model briefly. The EoW model consists of a JT black hole in AdS$_2$, a brane (the EoW brane) at the AdS boundary, and a coupled auxiliary system (see \mpref{eow}). The action of the whole system is \cite{replica1}
\begin{subequations}
\begin{align}
I_{\text{tot}} &= I_{\text{JT}} + \mu \int_{\text{brane}} d \ell, \label{total action1} \\
I_{\text{JT}}  &= - \frac{S_0}{2 \pi} \bigg[ \frac{1}{2} \int_{\mathcal{M}} \sqrt{g} R + \int_{\partial \mathcal{M}} \sqrt{h} \mathcal{K} \bigg] - \bigg[ \frac{1}{2} \int_{\mathcal{M}} \sqrt{g} \phi (R+2) + \int_{\partial \mathcal{M}} \sqrt{h} \phi \mathcal{K} \bigg], \label{jt action1}
\end{align}
\end{subequations}
where $\mu$ is the mass of brane and the integral is along the worldline of the brane. Here we choose the standard asymptotic boundary \cite{replica1}:
\begin{equation}
ds^2|_{\partial \mathcal{M}}= \frac{1}{\epsilon^2} d\tau^2, \qquad \phi=\frac{1}{\epsilon}, \qquad \epsilon \to 0,  \label{bondary condition1}
\end{equation}
where $\tau$ is the imaginary time of the boundary. For the EoW brane, the dual boundary condition is given by \cite{replica1}
\begin{equation}
\partial_{\vec{\text{n}}} =\mu \leq 0, \qquad \mathcal{K}=0, \label{eowbrane boundary condition}
\end{equation}
where $\partial_{\vec{\text{n}}}$ is the derivative normal to the boundary.

\par In order to capture the details of black hole evaporation, we focus attention on the case that the number of initial state $k$ in the EoW brane is very large. These states are used to describe the interior partners of the early Hawking radiation. More specifically, we consider the black hole system $B$ with an EoW brane. It has the number of dimensions $|B|=k$. Then we couple the black hole system $B$ to an auxiliary ``radiation'' system $R$  with $|R|=e^{S_0}$. We denote $\ket{\psi}$ as the whole pure system consisting of the black hole $B$ and radiation $R$. Therefore, the state of the whole system can be expressed as follows
\begin{equation}
\ket{\psi} = \frac{1}{\sqrt{k}} \sum_{i=1}^{k} \ket{\psi_i}_B \ket{i}_R, \label{system state }
\end{equation}
where $\ket{\psi_i}_B$ represents the black hole state, $\ket{i}_R$ represents the radiation. The index $i$ labels the entanglement between the black hole $B$ and the radiation $R$. The normalized factor is defined by
\begin{equation}
\braket{\psi_i| \psi_j}_B = \delta_{ij} Z_1,  \qquad \braket{i|j}_R = \delta_{ij}. \label{normalized factors}
\end{equation}
Here the notation $Z_n = Z_n (\beta)$ is the (replica) partition function on the AdS$_2$ disk topology with the boundary. This boundary consists of $n$ physical boundaries of renormalized length $\beta$ and $n$ EoW branes. We will use this notation frequently in the following replica geometry. Besides, as the evaporation proceeds, the black hole system $B$ radiates into the auxiliary system $R$. Then one can collect these Hawking particles in the auxiliary system. At this physical sense, we can consider $\log \big(\frac{k}{e^{S_0}}\big)$ as a time scale to characterize the stages of evaporation.

\par At present, we can trace out the black hole system $B$ to obtain the entanglement entropy of the radiation. We first need to find the (reduced) density matrix of radiation, which can be given by tracing out the black hole
\begin{equation}
\rho_R = \sum_l \braket{\psi_l | \Psi} \braket{\Psi | \psi_l} =\frac{1}{k} \sum_{i,j=1}^{k} \ket{j} \bra{i}_R \braket{\psi_i | \psi_j}_B. \label{density matrix1}
\end{equation}
Each element of the density matrix represents a gravitational amplitudes $\braket{\psi_i | \psi_j}$. It can be obtained by the following boundary condition
\begin{equation}
\braket{\psi_i | \psi_j} =
\begin{matrix}
 \includegraphics[width=5cm]{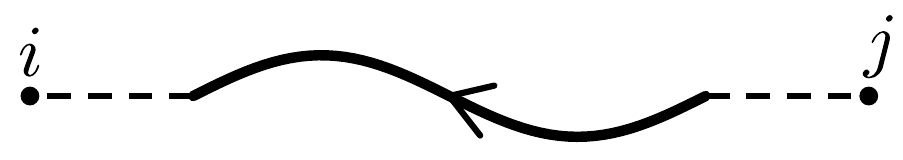}
\end{matrix}\,.   \label{grav amplitude}
\end{equation}
The solid line here is the asymptotic boundary that we impose in \eqref{bondary condition1}. The dashed line carries the index of state of EoW brane. The arrow direction indicates the direction of time. Then the entanglement entropy of radiation is given by
\begin{equation}
S_R \equiv - \text{Tr} (\rho_R \log \rho_R). \label{entanglement entropy of rad}
\end{equation}

\par However, in the actual calculation, if one uses the above formula \eqref{entanglement entropy of rad} directly to calculate, the process will be quite difficult because of the logarithmic term $\log \rho_R$ involved. A very clever mathematical trick is to first calculate the corresponding $n$-th R\'enyi entropy $S_n$, and then take the limit of $n \to 1$ to obtain the entanglement entropy
\begin{subequations}
\begin{align}
S_n &\equiv \frac{1}{1-n} \log \big [\text{Tr} (\rho_R^n) \big ], \label{renyi entropy2} \\
S_R &\equiv \lim\limits_{n \to 1}  \frac{1}{1-n} \log \big[ \text{Tr} (\rho_R^n) \big]. \label{von entropy}
\end{align}
\end{subequations}
We call this method  ``replica trick''. Namely, we make $n$ copies of the initial state $i$ and the final state $j$ and consider all possible path \eqref{grav amplitude}. Therefore, turn on the R\'enyi entropy for the auxiliary system $R$. The trace of the density matrix to the $n$-th power is expressed as follows
\begin{equation}
\text{Tr} (\rho_R^n) = \frac{1}{(k e^{S_0})^n} \sum_{i_1 \cdots i_n}^{k}  \braket{\psi_{i_1} | \psi_{i_2}}_B \cdot \braket{\psi_{i_2} | \psi_{i_3}}_B \cdots \braket{\psi_{i_n} | \psi_{i_1}}_B,  \label{trace density matrix}
\end{equation}
where we introduce the normalized factor $e^{n S_0}$, so $\text{Tr}(\rho_R) =1 $. Similar to \eqref{grav amplitude}, all multiplication of amplitudes $\prod \braket{\psi_{i_l} | \psi_{i_{l+1}} }_B$ are calculated by the gravitational path integral in the replica manifold $\mathcal{M}_n$. Its boundary are connected $n$-copies auxiliary system $R$. In the large $k$ limit, the replica wormholes saddle dominates the gravitational path integral, which leads to the planar topology approximation
\begin{equation}
\begin{split}
\text{Tr} (\rho_R^n) &\simeq \frac{1}{(kZ_1)^n} \big[ k (Z_1)^n + C_{n}^{2} k^2 Z_2 (Z_1)^{n-2} + \cdots + k^n Z_n \big] \\
                     &=\frac{1}{k^{n-1}} \bigg[ 1+ C_{n}^{2} \frac{kZ_2}{(Z_1)^2} + \cdots + \frac{k^{n-1}Z_n}{(Z_1)^n} \bigg], \label{planarapproximation}
\end{split}
\end{equation}
where $C_m^n$ is the combinatorial number, which indicates the number of different combinations of connected and/or disconnected geometries in the replica manifold. In fact, for the reality of black hole evaporation, all paths from the beginning of the initial state $\ket{\psi_i}$ to the end of the final state $\ket{\psi_j}$ should be taken into account. However, some paths are hard to find if we use the equation \eqref{entanglement entropy of rad}. When we use the expression \eqref{von entropy}, such paths will appear in non-trivial topologies and can thus be easily detected. This is another important reason for using the replica trick.

\par In the original work \cite{replica1}, the case $n=2$ is used as the most straightforward example\footnote{The simplest case of $n=2$ is enough for us to obtain the unitary result. Of course, we can also consider the contribution of higher order $n$, which appears in the sub-leading order \cite{subleading}.}. For this case, the physical quantity one will to calculate is called the ``purity''. In quantum information, it has an important property,
\begin{equation}
\begin{split}
\text{Tr} (\rho_R^2) &= \big[ \text{Tr} (\rho_R)  \big]^2,   \ \qquad \text{for the pure state}. \\
\text{Tr} (\rho_R^2) &\ll  \big[ \text{Tr} (\rho_R)  \big]^2, \qquad \text{for the mixed satae}.  \label{purity1}
\end{split}
\end{equation}
Thus, this relation allows us to determine whether the evaporation process is unitary. Now we consider the general case \eqref{trace density matrix}. As mentioned above, one should consider the summation of all possible topologies in the gravitational path integral, i.e., one should sum up \emph{all} possible internal connections. At its core, there are more than one summation ways to identify the geometry $\big | \braket{\psi_i | \psi_j}_B   \big |^n$. See \mpref{path} below.
\begin{figure}[htb]
\centering
\includegraphics[scale=0.55]{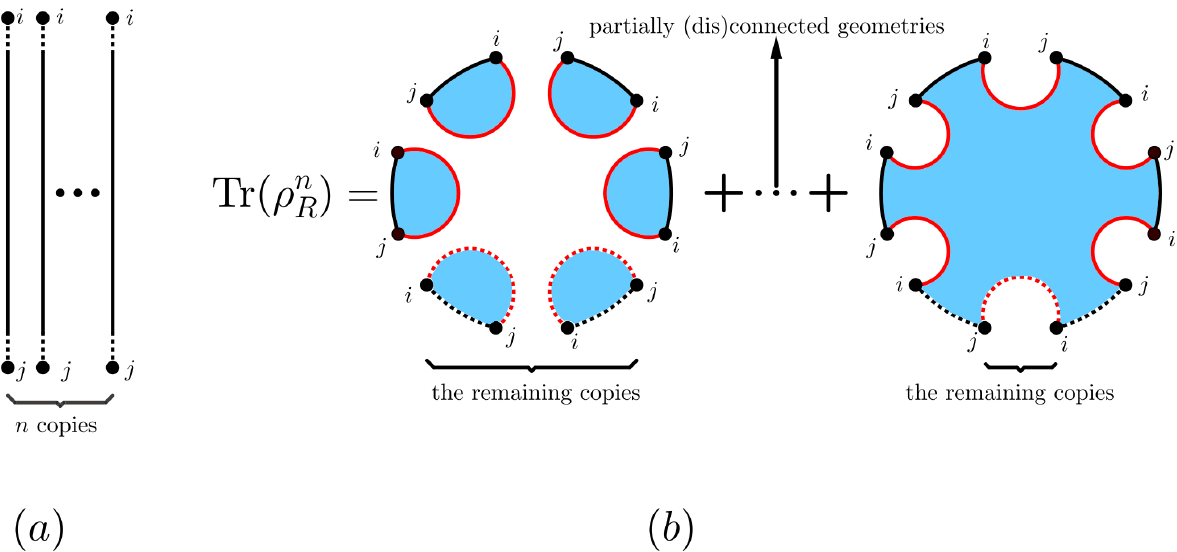}
\caption{\label{path} Schematic diagram for the gravitational path integral. The indices $i$ and $j$ represent different states. The gravitational region is shown in blue and the EoW brane is shown in red. (a) The boundary conditions for $\big | \braket{\psi_i | \psi_j}_B   \big |^n$. (b) The graphical representation of equation \eqref{trace density matrix}. Here we only plot two geometries that contribute in the leading order in the gravitational path integral. The dashed graphs represent omitted copies. The first geometry is the fully disconnected geometry. It is also called the ``Hawking saddle'', which dominates the evaporation at early times; The last geometry is the full connected geometry. It is also called the ``replica wormholes saddle'', which  dominates the evaporation at late times (after the Page time). The undrawn geometries are partially (dis)connected geometries. They can be ignored in the gravitational path integral as subleading terms and higher order terms. }
\end{figure}

\par For clarity and simplicity, we present only two classes of geometries (the full disconnected/connected geometry) in \mpref{path}. These correspond to the leading order term during the early and late stages of evaporation, respectively. Indeed, there are also partially (dis)connected geometries that contribute to the path integral. However, these contributions are considered as subleading terms and higher order terms in the whole evaporation and can therefore  be neglected.

\par It should be noted that the topological structures of these two geometries are entirely different! The former will reproduce Hawking's non-unitary calculations \cite{paradox}. We call it the ``Hawking saddle''. The latter connected geometry corresponds to some non-trivial paths in gravitational calculations. We call it the ``replica wormholes saddle''. The name comes from the fact that the interiors of black holes are connected and look like wormholes. Therefore, the equation \eqref{planarapproximation} obtained by the planar approximation is  expressed in the leading order as
\begin{equation}
\begin{split}
\text{Tr} (\rho_R^n) &\simeq \frac{k Z_1^2 + k^2 Z_2}{(kZ_1)^2} = \frac{1}{k} + \frac{Z_2}{Z_1^2} \\
                     &=k^{-1} + e^{-S_0}. \label{planarresult}
\end{split}
\end{equation}
The denominator in the first equation represents the normalization of the density matrix. The two terms in the second line represent the contributions of the connected geometry $(k^{-1})$ and the disconnected geometry $(e^{-S_0})$. Since $Z_n$ has the disk-like topology, in the planar approximation, we use the relation $Z_n \propto e^{S_0}$ to simplify. One can intuitively see that when $k$ is small enough (long before the Page time, since the time scale is set to $\log (k e^{-S_0})$), the disconnected geometry dominates, which reproduces the Hawking's result, i.e., a curve that grows linearly with time. However, when $k$ is very large (much later than the Page time), the connected geometry dominates, and the expression \eqref{planarresult} does not depend on $k$, which also suggests that the result does not increase with time at late times. It is the competition between these two saddles during the evaporation of black holes that leads to an unitary Page curve
\begin{equation}
\begin{split}
S_R             &\simeq \left \{
              \begin{array}{lr} \log k , \qquad \qquad \ \  \qquad t<t_{\text{Page}}. &\\
              \lim\limits_{n \to 1}S_{\text{BH}}^{(n)} =S_{\text{BH}}, \qquad t>t_{\text{Page}}. &
              \end{array} \right.  \label{rad entropy}
\end{split}
\end{equation}
where $S_{\text{BH}}^{(n)}$ is the R\'enyi black hole entropy: $S_{\text{BH}}^{(n)}= \frac{1}{1-n} \log \Big[ \frac{Z_n}{(Z_1)^n} \Big]$.

\par In the following, we provide the explicit calculation of the modular entropy and the CoE in the microcanonical ensemble and the canonical ensembles, respectively. The modular entropy yields to $n$-dependent Page curves (\mpref{modular microcanonical} and \mpref{modular canonical}) that differs from the original Page curve \cite{replica1,replica2}, while the CoE brings some new results (\mpref{coe microcanonical} and \mpref{coe canonical}).

\subsection{Replica Parameter Corrections to the Microcanonical Ensemble} \label{microcanonical}
\quad In the beginning, we need the explicit expression for the partition functions $Z_n(n=1,2 \cdots n)$ for the replica wormholes. The initial point is the boundary condition of $\text{Tr} (\rho_R^n)$. Namely, we need to sum over all replica geometries. For simplicity and computational convenience, we once again employ the planar approximation, where only the contributions from the planar geometry are considered to be significant, while those from non-planar geometries are exponentially suppressed  $(e^{-S_0})$ for large $S_0$. For the summation of planar geometries, it is accessible to use the resolvent matrix $R_{ij}(\lambda)$, where $\lambda$ refers to the eigenvalue for the reduced density matrix $\rho_R$. Now we define \cite{replica1}
\begin{subequations}
\begin{align}
R_{ij} (\lambda) &= \bigg( \frac{1}{\lambda \mathbb{1} -\rho_{R}}  \bigg)_{ij} = \frac{1}{\lambda} \delta_{ij} + \sum_{n=1}^{\infty} \frac{1}{\lambda^{n+1}} (\rho_R^n)_{ij}, \label{resolvent equation1} \\
R(\lambda) &= \sum_i R_{ij} (\lambda) = \text{Tr} [R_{ij}(\lambda)].  \label{resolvent equation2}
\end{align}
\end{subequations}
The entanglement spectrum of states can be derived by solving the Schwinger-Dyson (SD) equation for the resolvent matrix $R$. A detailed derivation of this calculation is provided in \cite{replica1}. So we do not delve into it extensively here. Finally, the density of state $D(\lambda)$ is determined by
\begin{equation}
D(\lambda) = \frac{1}{2 \pi i} \Big[ R(\lambda - i \epsilon) - R (\lambda + i \epsilon) \Big]. \label{density of state1}
\end{equation}
In terms of the density of state $D(\lambda)$, the R\'enyi entropy $S_n$ \eqref{renyi entropy1} and the modular entropy $S_{\text{mod}}$ \eqref{modular entropy1} are given by \cite{modular entropy}
\begin{subequations}
\begin{align}
S_n &= - \frac{1}{1-n} \log \bigg( \int_0^{\infty}  d\lambda \ \lambda^n D(\lambda) \bigg),   \label{sn1}  \\
S_{\text{mod}} &= S_n + n(n-1) \partial_n S_n \notag \\
               &= - \log \bigg( \int_0^{\infty} d\lambda \ \lambda^n D(\lambda) \bigg) - n \partial_n \Bigg[ \log \bigg( \int_0^{\infty} d\lambda \ \lambda^n D(\lambda) \bigg) \Bigg], \label{smod1}
\end{align}
\end{subequations}
and the CoE $C_n$ \eqref{entanglement capacity} is expressed as follows
\begin{equation}
\begin{split}
C_n &= -n \partial_n S_{\text{mod}}\\
&= n^2 \Bigg[ \frac{\int_0^{\infty} d\lambda \ \lambda^n (\log \lambda)^2 D(\lambda)}{\int_0^{\infty} d\lambda \ \lambda^n D(\lambda)} -  \bigg( \frac{\int_0^{\infty} d\lambda \ \lambda^n \log \lambda \ D(\lambda)}{\int_0^{\infty} d\lambda \ \lambda^n D(\lambda)}  \bigg)^2 \Bigg]. \label{cn1}
\end{split}
\end{equation}
In particular, we can obtain the expression for the von Neumann entropy and the quantum fluctuation (with respect to the original $\rho_R$) through take the limit of $n \to 1$
\begin{equation}
S_{\text{vN}} = - \int_0^{\infty} d\lambda \ D(\lambda) \ \lambda \log \lambda.  \label{s11}
\end{equation}
\begin{equation}
C_1= \int_0^{\infty} d\lambda \ D(\lambda) \ \lambda \ (\log \lambda)^2 - \bigg(\int_0^{\infty} d\lambda \ D(\lambda) \ \lambda \log \lambda \bigg)^2. \label{c11}
\end{equation}

\quad In the microcanonical ensemble, we respect the conservation of energy $E$ and fix $E= \frac{s^2}{2}$ in the asymptotic region rather than the renormalized length $\beta$. In this time the SD equation can be reduced to a quadratic equation about $R(\lambda)$ in the microcanonical ensemble \cite{replica1}. The corresponding solution determines the density of states $D(\lambda)$
\begin{subequations}
\begin{align}
R^2(\lambda) &+ \bigg( \frac{e^{S_0}-k}{\lambda} - k e^{S_0} \bigg) R(\lambda) + \frac{k^2 e^{S_0}}{\lambda} =0, \label{sd eq1}  \\
D(\lambda) &= \frac{k e^{S_0}}{2 \pi \lambda} \sqrt{\Big(\lambda - (k^{-\frac{1}{2}} - e^{-\frac{S_0}{2}})^2 \Big)  \Big( (k^{-\frac{1}{2}} - e^{-\frac{S_0}{2}}\big)^2 -\lambda   \Big)} + \delta(\lambda) (k-e^{S_0}) \theta(k-e^{S_0}) \notag \\
&= \frac{k^2}{2\pi \lambda^{\prime} t}  \sqrt{\Big( \lambda^{\prime} - (1-\sqrt{t})^2 \Big) \Big( (1+\sqrt{t})^2 -\lambda^{\prime} \Big)}+k^2 e^{-2S_0} \delta(\lambda^{\prime}) \Big(1-\frac{1}{t}\Big) \theta(t-1) \notag \\
&=k^2 D^{\prime} (\lambda^{\prime}), \label{sd eq2}
\end{align}
\end{subequations}
with the rescaled eigenvalues $\lambda^{\prime} \equiv k \lambda$, and we define the time scale factor $t= \frac{k}{e^{S_0}}$. When the Page time is reached, $t$ takes $1$. $e^{S_0}$ denotes as the number of states in a small interval near the energy $E$ (i.e., the energy bond of width $\Delta S$). The following normalization condition of the density $D(\lambda^{\prime})$ is determined by taking $\lambda^{\prime}=0$ in terms of the delta function:
\begin{subequations}
\begin{align}
\int^{\infty} d\lambda  D(\lambda) &= k \int^{\infty} d\lambda^{\prime}  D^{\prime} (\lambda^{\prime}) = k . \label{normalization1} \\
\int^{\infty} d\lambda \ \lambda D(\lambda) &= \int^{\infty}  d\lambda^{\prime} \ \lambda^{\prime} D^{\prime} (\lambda^{\prime}) =1. \label{normalization2}
\end{align}
\end{subequations}

\par Besides, due to the property of the step function $\theta(1-t)$, we have:\\
(I)\ When $0<t<1$ (before the Page time), there are $k$ states distributed over the interval $\lambda^{\prime} \in A = \big[(1-\sqrt{t})^2, (1+\sqrt{t})^2 \big]$. In this case, the density of states is a real value.\\
(II)\ When $t>1$ (after the Page time), there are $e^{S_0}$ states located at the same interval, and $(k-e^{S_0})$ states located at $\lambda^{\prime}=0$.

\par
\begin{figure}[htb]
\centering
\subfigure[\scriptsize{}]{\label{mod1}
\includegraphics[scale=0.55]{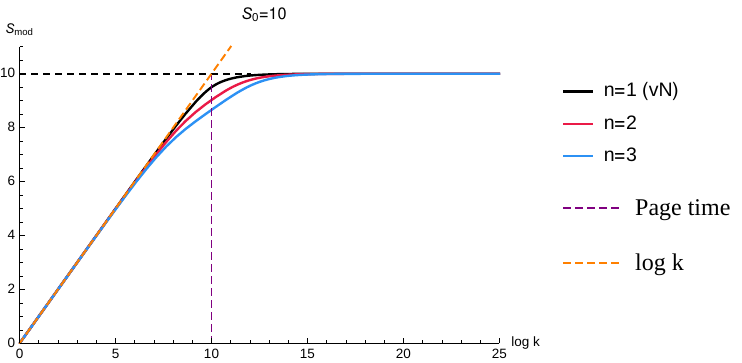}
}
\quad
\subfigure[\scriptsize{}]{\label{mod2}
\includegraphics[scale=0.4]{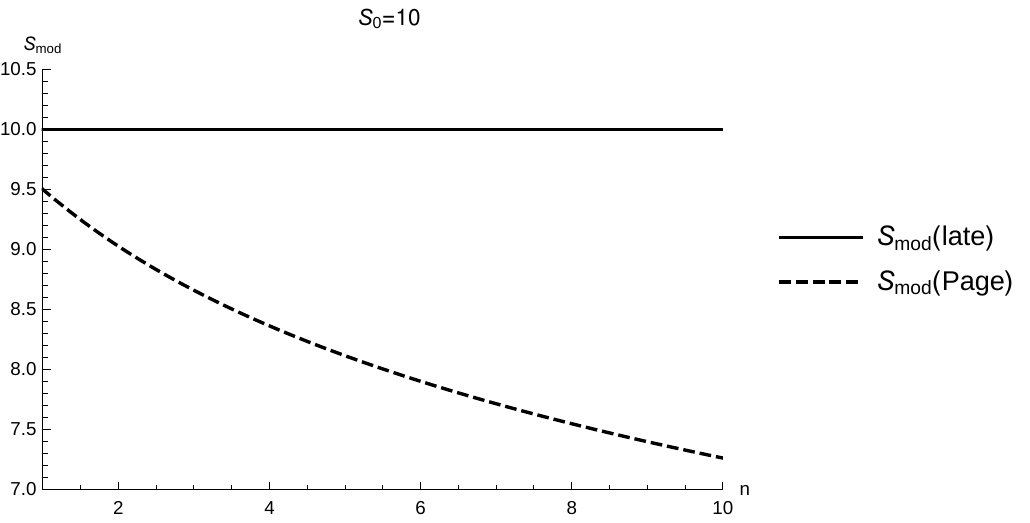}
}
\caption{(a) The modular entropy for the EoW model with $S_0=10$ in the microcanonical ensemble. In order to evaluate the evolution more intuitively, here we rescale the time coordinate $t$ in terms of $\log k$ by the relation $t= \frac{ k}{e^{S_0}}$. The Page time is $\log k=S_0=10$. At early times, the entropy approaches a thermal state and increases as $\log k$. After the Page time, its value is approximately a constant $S_0=10$. Contrary to common sense, the entropy is not at the maximum at the Page time. Moreover, the transition between the two phases is smooth at this moment. (b) The value of modular entropy at the Page time changes with the replica parameter $n$. The larger the replica parameter $n$, the smaller the value of modular entropy at the Page time.}
\label{modular microcanonical}
\end{figure}

\begin{figure}[htb]
\centering
\subfigure[\scriptsize{}]{\label{coe1}
\includegraphics[scale=0.5]{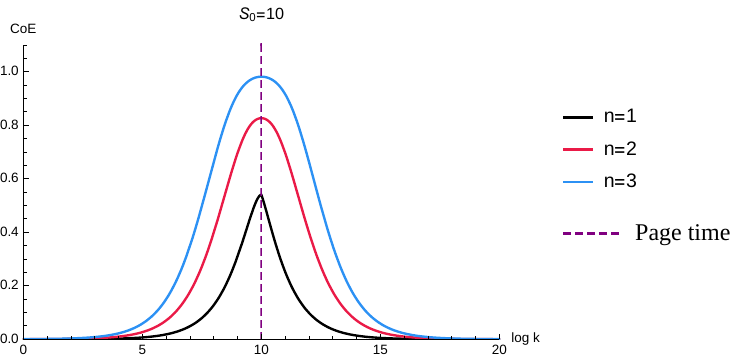}
}
\quad
\subfigure[\scriptsize{}]{\label{coe2}
\includegraphics[scale=0.35]{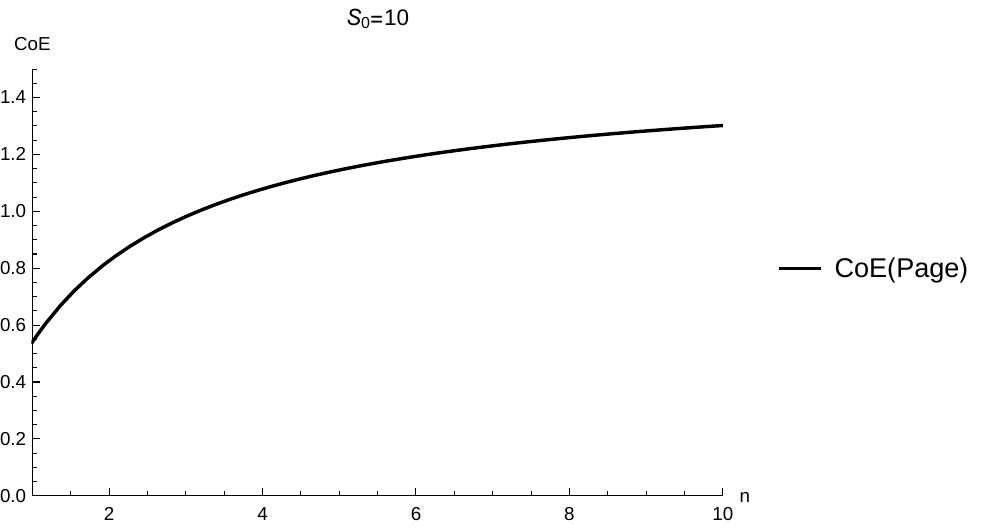}
}
\caption{(a) The CoE with $S_0=10$ for the microcanonical ensemble. The CoE increases from zero until the Page time ($\log k=10$) reaches a maximum, then decreases to zero at late times. The greater $n$, the faster the CoE increases, while the longer it takes for the corresponding curve to approach zero. (b) The value of CoE at the Page time varies with $n$. Its maximum value increases as $n$ increases.}
\label{coe microcanonical}
\end{figure}

 In fact, it has been shown by Page's theorem \cite{page theorem} that in the planar limit, the density of state $D^{\prime}(\lambda^{\prime})$ is just the entanglement spectrum of a subsystem of dimension $k$ in a random state of the total system spanned by Hilbert space of size $k e^{S_0}$. Therefore, we can deduce the expressions of the modular entropy \eqref{smod1} and the CoE \eqref{cn1} by integrating $\lambda^{\prime}$ as follows:
\begin{subequations}
\begin{align}
S_n            &=S_0 + \log t +\frac{1}{1-n} \log \bigg( \int_A d\lambda^{\prime} \ \lambda^{\prime n} D^{\prime} (\lambda^{\prime}) \bigg), \label{sn2} \\
S_{\text{mod}} &=S_n+n(n-1)\partial_n S_n   \notag \\
               &= S_0 +\log t - \log \bigg( \int_A  d\lambda^{\prime} \ \lambda^{\prime n} D^{\prime} (\lambda^{\prime}) \bigg)- n\partial_n \Bigg[ \log \bigg( \int_A d\lambda^{\prime} \ \lambda^{\prime n} D^{\prime} (\lambda^{\prime}) \bigg) \Bigg],  \label{smod2}
\end{align}
\end{subequations}
and
\begin{equation}
\begin{split}
C_n &= -n \partial_n S_{\text{mod}}   \\
               &= n^2 \Bigg[ \frac{\int_A d\lambda^{\prime} \ \lambda^{\prime n}  (\log \lambda^{\prime})^2 D^{\prime} (\lambda^{\prime})}{\int_A d\lambda^{\prime} \ \lambda^{\prime n} D^{\prime} (\lambda^{\prime})} - \Bigg( \frac{\int_A d\lambda^{\prime} \ \lambda^{\prime n}  \log \lambda^{\prime} \ D^{\prime} (\lambda^{\prime}) }{\int_A d\lambda^{\prime} \ \lambda^{\prime n} D^{\prime} (\lambda^{\prime}} \Bigg)^2 \Bigg]. \label{cn2}
\end{split}
\end{equation}
In the above equations, we first evaluate the following integral
\begin{equation}
\begin{split}
\tilde{I} = \int_A \lambda^{\prime n} D^{\prime} (\lambda^{\prime}) d\lambda^{\prime} &= \int_{(1-\sqrt{t})^2}^{(1+\sqrt{t})^2} \frac{1}{2 \pi \lambda^{\prime} t} \sqrt{\Big[ \lambda^{\prime}-(1-\sqrt{t})^2 \Big] \Big[ (1+\sqrt{t})^2-\lambda^{\prime}\Big]} \ \lambda^{\prime} \ D^{\prime} (\lambda^{\prime}) d\lambda^{\prime} \\
&=\frac{2^{2n+2} (\sqrt{t})^{n-1}}{\pi} \int_0^1 \sqrt{(1-\Lambda) \Lambda} \bigg(  \Lambda+\frac{(1-\sqrt{t})^2}{4\sqrt{t}} \bigg)^{n-1} d\Lambda \\
&=(1-\sqrt{t})^{2n-2} \  _2F_1\bigg(1-n,\frac{3}{2},3,\frac{-4\sqrt{t}}{(1-\sqrt{t})^2}\bigg) \\
&={_2}F_1 \bigg(1-n,-n, 2,t \bigg) = t^{n-1} {_2F_1}\bigg(1-n,-n, 2,\frac{1}{t} \bigg),  \label{integral1}
\end{split}
\end{equation}
where we rescale the parameter $\Lambda$ by $\Lambda = \frac{\sqrt{t} \lambda^{\prime} - (1-\sqrt{t})^2 \sqrt{t}}{4t}$, and use the Euler transformation and the Kummer transformation for hypergeometric function $_2F_1(a,b,c,d)$ in the last two lines. Note that, based on two points (I) and (II) above, this expression is valid only before the Page time ($t \in [0,1]$). However, one can simply replace $t$ with $\frac{1}{t}$ to obtain the similar expression after the Page time, namely
\begin{equation}
\begin{split}
\tilde{I}            &= \left \{
              \begin{array}{lr} \tilde{I}_1=\ _2F_1(1-n,-n,2,t) , \qquad \   0\le t \le 1. &\\
              \tilde{I}_2=t^{n-1} \ _2F_1\big(1-n,-n,2,\frac{1}{t}\big),  \qquad   t \ge 1. &
              \end{array} \right.  \label{integral2}
\end{split}
\end{equation}
Then, we finally obtain the R\'enyi entropy and the modular entropy as follows\footnote{For $n=1$ case, we need to use some approximation relations to obtain the explicit expression, see appendix \ref{appendixa}.}
\begin{equation}
\begin{split}
S_n  =S_0 + \log t +\frac{1}{1-n} \log \tilde{I}         &= \left \{
              \begin{array}{lr} S_0 + \log t + \frac{1}{1-n} \log \tilde{I}_1, \qquad \   0\le t \le 1. &\\
              S_0 + \log t + \frac{1}{1-n} \log \tilde{I}_2,  \qquad  \qquad \  t \ge 1. &
              \end{array} \right.  \label{sn microcanonical}
\end{split}
\end{equation}
and
\begin{equation}
\begin{split}
S_{\text{mod}}  =S_n + n(n-1)\partial_n S_n        &= \left \{
              \begin{array}{lr} S_0 + \log t +  \log \tilde{I}_1 - n \frac{\partial_n \tilde{I}_1}{\tilde{I}_1}, \qquad \   0\le t \le 1. &\\
              S_0 + \log t +  \log \tilde{I}_2 - n \frac{\partial_n \tilde{I}_2}{\tilde{I}_2},  \qquad  \qquad \  t \ge 1. &
              \end{array} \right.  \label{smod microcanonical}
\end{split}
\end{equation}
Besides, the CoE can also be obtained by taking the derivative of the modular entropy
\begin{equation}
\begin{split}
C_n  =-n \partial_n S_{\text{mod}}       &= \left \{
              \begin{array}{lr} \frac{n^2 (\tilde{I}_1 \cdot \partial_n^2 \tilde{I}_1 - (\partial_n \tilde{I}_1)^2 )}{\tilde{I}_1^2}, \qquad \   0\le t \le 1. &\\
              \frac{n^2 (\tilde{I}_2 \cdot \partial_n^2 \tilde{I}_2 - (\partial_n \tilde{I}_2)^2 )}{\tilde{I}_2^2},  \qquad  \qquad \  t \ge 1. &
              \end{array} \right.  \label{cn microcanonical}
\end{split}
\end{equation}
At last, we can plot the corresponding curves for the modular entropy and the CoE in \mpref{modular microcanonical} and \mpref{coe microcanonical}, respectively. Especially, the case of the von Neumann entropy for the limit $n \to 1$ is
\begin{equation}
\begin{split}
S_{\text{vN}}  =\lim\limits_{n \to 1} S_n (\text{or} \ S_{\text{mod}})      &= \left \{
              \begin{array}{lr} S_0 + \log t = \log k, \qquad \   t \gg 1. &\\
              S_0,  \qquad  \qquad \qquad \qquad \ \  t \ll 1. &
              \end{array} \right.  \label{vn microcanonical}
\end{split}
\end{equation}
which also follows our previous calcluation \eqref{rad entropy} for the behavior of entropy of radiation.

\par However, for the general replica parameter $n$, we find that, for the modular entropy, each curve grows approximately linear at early times, then reaches to the saturation value $S_0=10$ at late times (\mpref{mod1}). When the replica parameter $n$ increases, the entropy increases slowly, and the corresponding values at the Page time decrease (\mpref{mod2}). For the CoE, each case is a curve that starts from zero and increases, then decreases to zero eventually (\mpref{coe1}). They all reach their maximum at the Page time. The maximum value increases as $n$ increases (\mpref{coe2}). It is important to note that at late times, both modular entropy and CoE become independent on the replica parameter $n$. This contrasts significantly with the subsequent case of canonical ensembles. In addition, since the CoE is defined by the derivative of modular entropy with respect to $n$ \eqref{entanglement capacity}. When the modular entropy becomes a constant that is independent of $n$ at late times, the CoE  decays to zero (\mpref{coe1}). As we will see later, the CoE in the canonical ensemble is different from this result, as it approximates the heat capacity and does not equal zero at late times.

\subsection{Replica Parameter Corrections to the Canonical Ensemble}  \label{canonical}
\begin{figure}[htb]
\centering
\subfigure[\scriptsize{}]{\label{mod3}
\includegraphics[scale=0.4]{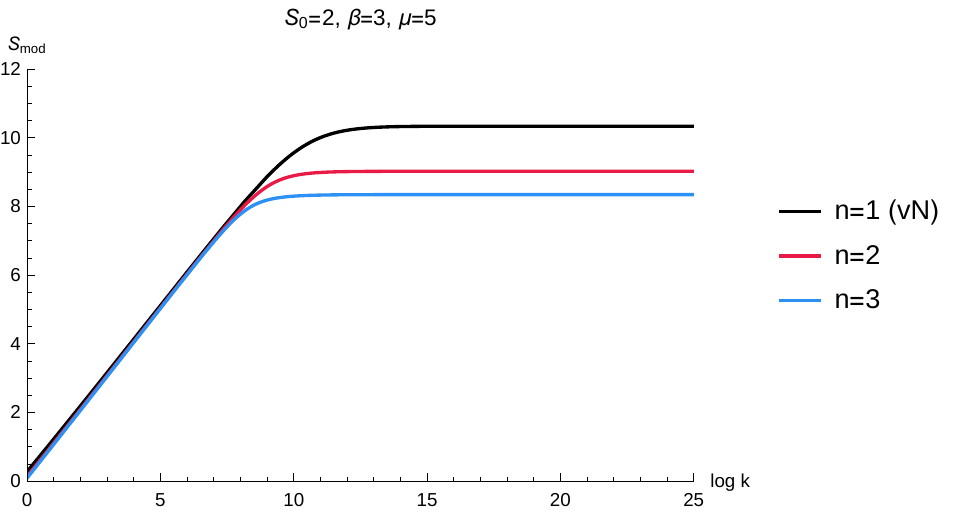}
}
\quad
\subfigure[\scriptsize{}]{\label{mod4}
\includegraphics[scale=0.4]{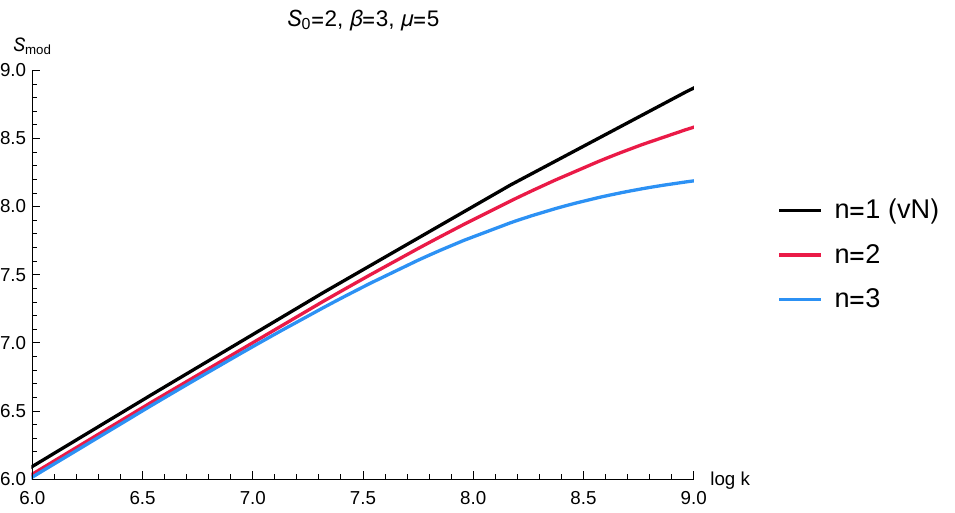}
}
\quad
\subfigure[\scriptsize{}]{\label{mod5}
\includegraphics[scale=0.4]{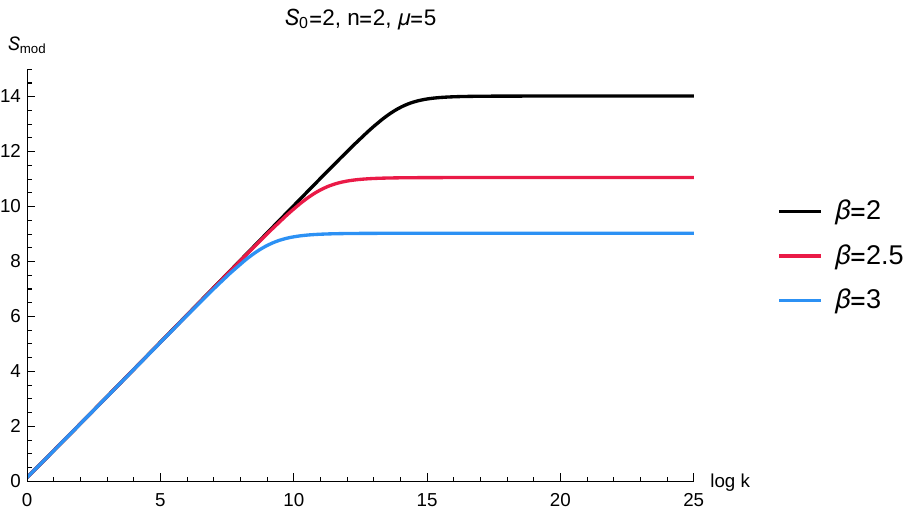}
}
\quad
\subfigure[\scriptsize{}]{\label{mod6}
\includegraphics[scale=0.4]{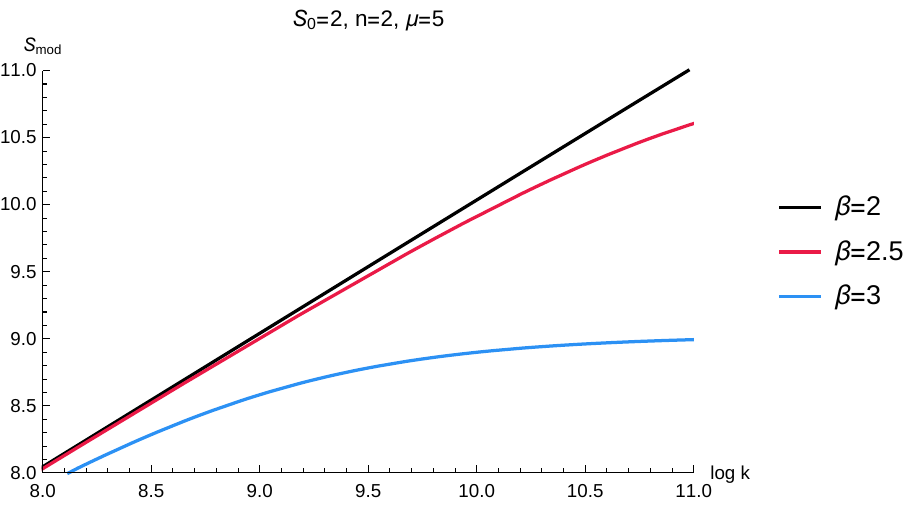}
}
\caption{The modular entropy for the EoW model in the canonical ensemble. Here we convert the time scale to $\log k=S_0+ \log \bigg( \frac{2\pi s_k \cosh(2\pi s_k) - \sinh (2\pi s_k)}{8\pi^4} \bigg)$ \eqref{definition of k}. (a) and its zoomed plot (b) illustrate the evolution curves of modular entropy with fixed $\beta=3$ as a function of the replica parameter $n$, whereas (c) and its zoomed plot (d) depict the evolution curves of modular entropy with fixed $n=2$ with respect to the inverse temperature $\beta$. Each curve exhibits an approximately linear increase before approaching a saturation value. The influence of parameter $n$ on the modular entropy is analogous to that of the inverse temperature $\beta$. Specifically, the final saturation value decreases as either $n$ or $\beta$ increases.}
\label{modular canonical}
\end{figure}

\begin{figure}[htb]
\centering
\subfigure[\scriptsize{}]{\label{coe3}
\includegraphics[scale=0.4]{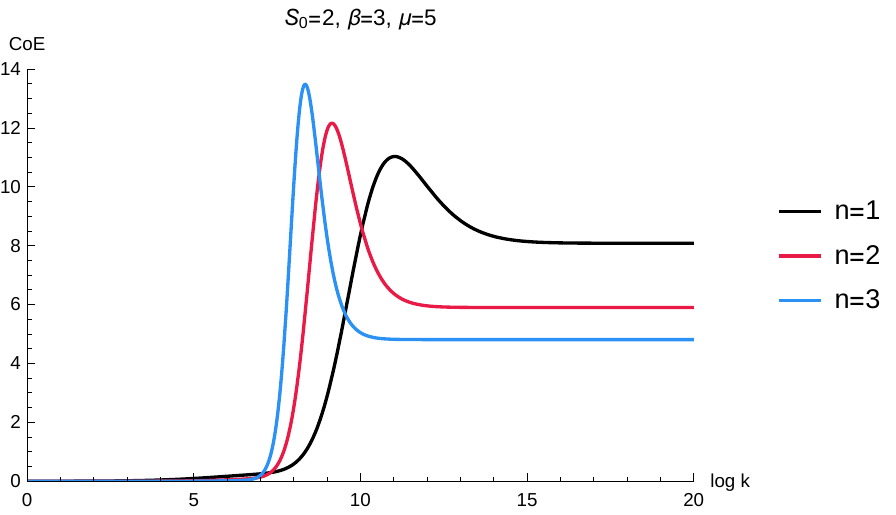}
}
\quad
\subfigure[\scriptsize{}]{\label{coe4}
\includegraphics[scale=0.4]{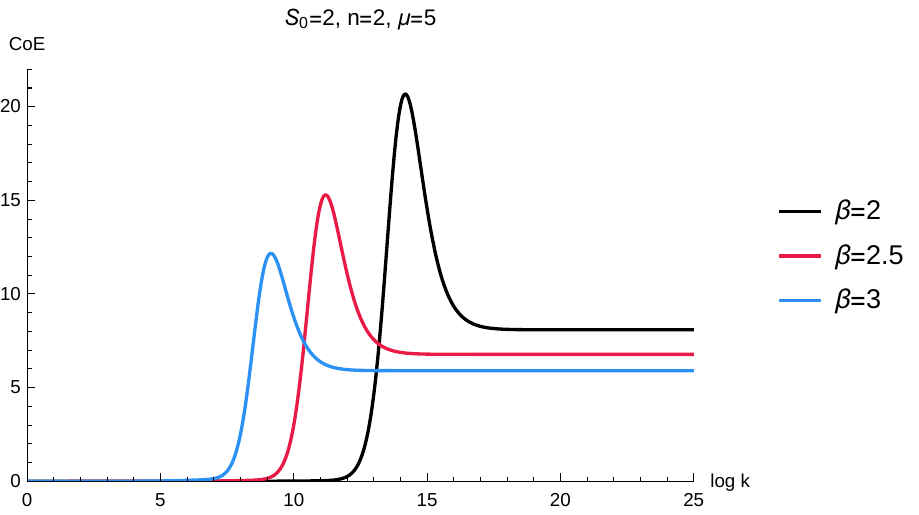}
}
\caption{The CoE in the canonical ensemble. Each curve grows from zero, reaches its peak at the Page time, then it decline and reaches a constant. (a) The CoE with fixed $\beta=3$ changes with the replica parameter $n$. (b) The CoE with fixed $n=2$ as a function of the inverse temperature $\beta$. Similar to the modular entropy, the replica parameter $n$ and the inverse temperature $\beta$ have similar effects on the saturating value at late times. The larger $n$ or $\beta$, the smaller the saturating value. }
\label{coe canonical}
\end{figure}

\quad Next, we proceed to the canonical ensemble. In this case, the temperature $\frac{1}{\beta}$ is fixed instead of the energy $E= \frac{s^2}{2}$. Therefore, this leads to an outcome that differs from in the case of microcanonical ensembles (see \mpref{modular canonical} and \mpref{coe canonical}). Especially the behavior of the modular entropy and the CoE at late times.  We will try to calculate the resolvent equation \eqref{resolvent equation1} numerically by following \cite{replica1}, since the density of state can not be analytically solved in the canonical ensemble. For the purpose of calculating the modular entropy and CoE in the semi-classical background. We focus on the small $G_N$ regime. Besides, this parameter is not explicitly expressed in the EoW model. In fact, it can be recovered by the inverse temperature through $\beta= \beta G_N$. Therefore, the small $G_N$ limit is equivalent to the small $\beta$ limit, which is the high temperature limit. We will take this limit in the following content.
\par In the beginning, we solve the SD equation in the planar approximation to obtain the resolvent $R(\lambda)$
\begin{equation}
\lambda=\frac{k}{R(\lambda)} + \int_{0}^{\infty} ds \tilde{\rho}(s) \frac{w(s)R(\lambda)}{k-w(s)R(\lambda)}, \label{sd eq3}
\end{equation}
where the parameter are defined as follows:
\begin{subequations}
\begin{align}
        \tilde{\rho}(s) &= e^{S_0} \rho(s) = e^{S_0} \frac{s}{2\pi^2} \sinh (2 \pi s),  \label{parameter rho} \\
             Z_n&= \int_{0}^{\infty} ds   \tilde{\rho}(s) y^n(s), \label{parameter zn} \\
            y(s)&= e^{-\frac{\beta s^2}{2}} 2^{1-2\mu} \Big|\Gamma \Big(\mu - \frac{1}{2}+is \Big) \Big|^2, \label{parameter ys} \\
            w(s)&= \frac{y(s)}{Z_1}. \label{parameter ws}
\end{align}
\end{subequations}
Here $Z_n$ is the replica partition function, $\mu$ is the brane mass. The planar limit is refer to $k$ or $e^{S_0}= \frac{k}{t}$ is large. Before proceeding, we explain the physical properties of entanglement spectrum in the canonical ensemble. The evolution of the density of state $D(\lambda)$ in canonical ensembles is mainly divided into two stages. Before the Page transition, $D(\lambda)$ is concentrated near $\lambda=\frac{1}{k}$ at early times. However, at and after the Page transition, the distribution gradually approaches the thermal spectrum, which produces a sharp cutoff at $s=s_k$, namely
\begin{equation}
k \equiv  \int_{0}^{s_k} ds \ \tilde{\rho}(s) = e^{S_0}  \cdot \frac{2\pi s_k \cosh (2\pi s_k)-\sinh(2\pi s_k)}{8 \pi^4},  \label{definition of k}
\end{equation}
Then, we can acquire the minimal eigenvalue $\lambda_0$ of $D(\lambda)$, which is corresponds to a location that $\frac{d\lambda}{dR}=0$. By the equation \eqref{sd eq3}, we have
\begin{subequations}
\begin{align}
  \lambda_0 &\simeq \frac{1}{k} \int_{s_k}^{\infty} ds \tilde{\rho}(s) w(s) \notag \\
            &=\frac{1}{k} \bigg( 1- \int_{0}^{s_k} ds  \tilde{\rho}(s) w(s) \bigg). \label{minimal eigenvalue1} \\
R(\lambda_0)&\simeq -\frac{k}{w(s_k)}. \label{minimal eigenvalue2}
\end{align}
\end{subequations}
To the next, we split the integral into two parts
\begin{equation}
\lambda R \simeq k+ \int_{0}^{s_k} ds \tilde{\rho}(s) \frac{w(s)R}{k-w(s)R} + \frac{R}{k} \int_{s_k}^{\infty} ds \tilde{\rho}(s) w(s). \label{sd eq4}
\end{equation}
Qualitatively speaking, these two parts correspond to the before/after the Page transition, respectively\footnote{The reasons for the feasibility of equation \eqref{sd eq4} are detailed in \cite{replica1}.}. The last term corresponds to the high energy state, which can be approximated to $\lambda_0 R$. Thus
\begin{equation}
(\lambda-\lambda_0)R \simeq k +\int_{0}^{s_k} \tilde{\rho}(s) \frac{w(s)R}{k-w(s)R} ds. \label{sd eq5}
\end{equation}
The second term can be regards as a small perturbation, resulting in a first-order solution
\begin{equation}
R(\lambda)\simeq \frac{k}{\lambda-\lambda_0} + \frac{1}{\lambda-\lambda_0} \int_{0}^{s_k} \tilde{\rho}(s) \frac{w(s)}{\lambda-\lambda_0-w(s)}. \label{sd eq6}
\end{equation}
Under the following condition
\begin{equation}
\begin{split}
                 k &\gg \int_{0}^{s_k} \tilde{\rho}(s) \frac{w(s)}{\lambda-\lambda_0-w(s)}, \\
\text{or} \ \lambda &> \lambda_0+w(s_k-\delta),
\end{split}
\end{equation}
with a control parameter $\delta$. In the end, we finally obtain the density of state is expressed as \cite{replica1}
\begin{equation}
D(\lambda) = \int_0^{s_k} \tilde{\rho}(s) \delta(\lambda-\lambda_0-w(s)), \label{density of state2}
\end{equation}
with the following two normalization conditions
\begin{subequations}
\begin{align}
\int_0^{\infty} d\lambda \ D(\lambda) &= k, \label{normalization3} \\
\int_0^{\infty} d\lambda \ \lambda \ D(\lambda)   &=1. \label{normalization4}
\end{align}
\end{subequations}
Though this result, further, we invoke the result \eqref{density of state2} into expressions \eqref{sn1}, \eqref{smod1} and \eqref{cn1}, and find that the entropy is written as
\begin{subequations}
\begin{align}
S_n &= \frac{1}{1-n} \log \bigg( \int_0^{s_k} ds \tilde{\rho}(s) (\lambda_0+w(s))^n \bigg), \label{sn3}  \\
S_{\text{mod}} &= - \log \bigg( \int_0^{s_k} ds \tilde{\rho}(s) (\lambda_0+w(s))^n \bigg) \notag \\
               &- n \partial_n \bigg[ \log \bigg( \int_0^{s_k} ds \tilde{\rho}(s) (\lambda_0+w(s))^n \bigg) \bigg], \label{smod3}
\end{align}
\end{subequations}
and the CoE is expressed by
\begin{equation}
\begin{split}
C_n &= n^2 \Bigg[ \frac{\int_0^{s_k} ds \tilde{\rho}(s) (\lambda_0+w(s))^n \big( \log (\lambda_0+w(s))\big)^2}{\int_0^{s_k} ds \tilde{\rho}(s) (\lambda_0+w(s))^n}  \\
&- \Bigg( \frac{\int_0^{s_k} ds \tilde{\rho}(s) (\lambda_0+w(s))^n \log (\lambda_0+w(s))}{\int_0^{s_k} ds \tilde{\rho}(s) (\lambda_0+w(s))^n} \Bigg) \Bigg]. \label{cn3}
\end{split}
\end{equation}
In the limit of $n \to 1$, we have
\begin{equation}
S_{\text{vN}} = - \int_{0}^{s_k} ds \tilde{\rho}(s) (\lambda_0+w(s)) \log (\lambda_0 +w(s)). \label{s12}
\end{equation}
and
\begin{equation}
\begin{split}
C_1 &= \int_{0}^{s_k} ds \tilde{\rho}(s) (\lambda_0+w(s)) \big(\log (\lambda_0 +w(s)) \big)^2 \\
&-\bigg(\int_{0}^{s_k} ds \tilde{\rho}(s) (\lambda_0+w(s)) \log(\lambda_0+w(s)) \bigg)^2. \label{c12}
\end{split}
\end{equation}
\par Through the numerical calculation, we plotted the corresponding images in \mpref{modular canonical} (for the modular entropy as the function of the replica parameter $n$ and the inverse temperature $\beta$) and  \mpref{coe canonical} (for the CoE changes with $n$ and $\beta$) by setting $S_0=2, \mu=5$.

\par We now provide some analytical calculations to validate our numerical results \eqref{smod3}, \eqref{cn3}, \eqref{s12}, and \eqref{c12}. For simplicity, we concentrate on the behavior of the relevant physical quantities at early and late times. At early times, namely, in the small $k$ region, the density of state $D(\lambda)$ is sharply peaked around $\frac{1}{k}$ \cite{replica1}
\begin{equation}
D(\lambda)(\text{early}) \simeq \int_{0}^{s_k} ds \tilde{\rho}(s) \delta \Big(\lambda-\frac{1}{k}\Big). \label{density of state early}
\end{equation}
So the R\'enyi entropy behaves as
\begin{equation}
\begin{split}
S_n^{\text{(early)}} &=\frac{1}{1-n} \log \bigg[  \int_0^{\infty} D(\lambda)(\text{early})\ \lambda^n \ d\lambda  \bigg] \\
                   &\simeq \frac{1}{1-n} \log \bigg[  \int_0^{\infty} D(\lambda)(\text{early})  \bigg( \frac{1}{k} + \Big(\lambda-\frac{1}{k}\Big) \bigg)^n d \lambda \bigg] \\
                   & \simeq \log k - \frac{nk}{2} \frac{Z_2}{Z_1^2} \sim \log k - \mathcal{O}(10^{-6}) \cdot nk, \label{sn canonical early}
\end{split}
\end{equation}
with $Z_1 \simeq 107.98$ and $Z_2 \simeq 0.38$ \eqref{parameter zn} for $\beta=3$. In the first approximately equal sign, we expand $\lambda$ about $\frac{1}{k}$ to the first order. In the second approximately equal sign, we use the following relations
\begin{equation}
\int d \lambda D(\lambda) =0, \qquad \int d\lambda D(\lambda) \Big( \lambda - \frac{1}{k} \Big)=0, \qquad \int d\lambda D(\lambda) \Big(\lambda - \frac{1}{k} \Big)^n = n \frac{Z_2}{Z_1^2}.
\end{equation}
Then the modular entropy and the CoE at early times are
\begin{equation}
S_{\text{mod}}^{\text{(early)}}=S_n^{\text{(early)}}+n(n-1) \partial_n S_n^{\text{(early)}} \simeq \log k -  \mathcal{O}(10^{-7}) \cdot n^2 k. \label{smod canconical early}
\end{equation}
and
\begin{equation}
C_n^{\text{(early)}} = -n \partial_n S_{\text{mod}}^{\text{(early)}} \simeq \mathcal{O}(10^{-4}) \cdot n^2 k.  \label{cn canonical early}
\end{equation}
This conforms to the linear behavior of the modular entropy (\mpref{mod4} and \mpref{mod6}) as well as the exponential behavior of the CoE (\mpref{coe3} and \mpref{coe4}) at early times (note that we are use $\log k$ as the variable).

\par Next, we regard the large $k$ region as the later period. The state density distribution is very close to a complete thermal spectrum. We can use the approximation as follows
\begin{equation}
\lambda_0 \sim \frac{1}{k} \simeq 0, \qquad D(\lambda)(\text{late}) \simeq \int_0^{s_k} \tilde{\rho}(s) \delta(\lambda - w(s)). \label{density of state late}
\end{equation}
In fact, in this scenario, the R\'enyi entropy approaches to the coarse-grained or R\'enyi black hole entropy $S_{\text{BH}}^{(n)}$
\begin{equation}
\begin{split}
S_n^{\text{(late)}} \simeq S_{\text{BH}}^{(n)} &=\frac{1}{1-n} \log \bigg( \int_0^{\infty} D(\lambda)(\text{late})  \lambda^n \ d\lambda \bigg) \simeq \frac{1}{1-n} \log \bigg( \int_0^{s_k}  w^n(s) \tilde{\rho}(s) ds \bigg) \\
& \simeq \frac{1}{1-n}  \bigg( \int_0^{s_k} y^n(s) \tilde{\rho}(s) ds \bigg) -\frac{n}{1-n} \log Z_1.
\end{split}
\end{equation}
We assume that the brane mass is large, i.e., $\mu \gg \frac{1}{\beta}$ for simplicity. Taking the following relation
\begin{subequations}
\begin{align}
\frac{\Big| \Gamma \Big(\mu -\frac{1}{2} +is \Big) \Big|^2}{\Big | \Gamma \Big(\mu -\frac{1}{2}\Big) \Big|^2} & \sim 1, \qquad y^n(s) \sim y(0) e^{-\frac{n\beta}{2}s^2}.  \\
Z_1 &= \int_0^{\infty} \tilde{\rho}(s) y(s) ds \sim e^{S_0} \frac{e^{\frac{2\pi^2}{\beta}}}{\beta^{\frac{3}{2}}}.
\end{align}
\end{subequations}
Then, we obtain the R\'enyi black hole entropy as follows
\begin{equation}
S_n^{\text{(late)}} \simeq S_{\text{BH}}^{(n)}=S_0 + \frac{2(1+n) \pi^2}{n \beta } + {\cal O} \bigg( \log \frac{2\pi}{n\beta} \bigg). \label{sn canonical late}
\end{equation}
So we can also obtain the modular entropy and the CoE at late times are given by
\begin{equation}
S_{\text{mod}}^{\text{(late)}} \simeq S_{\text{BH}}^{(n)} + n(n-1) \partial_n S_{\text{BH}}^{(n)} =S_0 + \frac{4\pi^2}{n\beta } \propto \frac{1}{n \beta}, \label{smod canonical late}
\end{equation}
and
\begin{equation}
C_n^{\text{(late)}} = -n \partial_n S_{\text{mod}}^{\text{(late)}} \simeq \frac{4\pi^2}{n \beta} \propto \frac{1}{n \beta} \neq0. \label{cn canonical late}
\end{equation}
From the equations \eqref{smod canonical late} and  \eqref{cn canonical late}, it is evident that the replica parameter $n$ and the inverse temperature $\beta$ are coupled in the denominator. This relation also explains the decrease in the physical quantity as either $n$ or $\beta$ increases at late times.
On the one hand, the $n$-dependent results indicate that the greater number of wormholes in the later stage, the more Hawking radiation is purified, and the fully connected geometry is increasingly dominant.  On the other hand, the proportional relationship with $\frac{1}{n \beta}$ also provides the similar impacts of the replica parameter and the inverse temperature on both modular entropy and CoE (see \mpref{mod3}, \mpref{mod5}, and \mpref{coe canonical}). This interdependence also suggests a profound relationship between $n$ and $\beta$. We will thoroughly investigate these points in the subsequent section. Therefore, we verify the rationality of \mpref{modular canonical} and \mpref{coe canonical} through some reasonable approximations at early and late times.

\par In brief, both the modular entropy and the CoE in the canonical ensemble approach saturating values at late times. However, this result differs from that obtained in the microcanonical ensemble.  These saturation values decrease as the $n$ or $\beta$. In particular, the CoE remains non-zero at late times, which is in contrast to the case for the microcanical ensemble \eqref{cn microcanonical}. Because for the canonical ensemble with fixed $\beta$, the CoE actually approximates the heat capacity that defined by $\frac{1}{\beta}$ at late times. It remains a constant and does not decay to zero. If we consider an evaporating black hole with a varying $\beta$, its modular entropy and CoE will not remain at a saturated value. Finally, combining our previous findings in the microcanonical ensemble, we provide a comprehensive summary of the behavior of the modular entropy and the CoE for the EOW model in Table \ref{table2}.

\begin{table}[htb!]
\centering
 \fontsize{10}{14}\selectfont
\setlength{\tabcolsep}{1mm}
   \begin{tabular}{|c|c|c|c|}
    \hline
\thead{\diagbox{\textbf{Quantity}}{\textbf{Stage}}} &
\multicolumn{2}{|c|}{\textbf{Before the Page transition}} &
  \thead{\textbf{After the Page transition}} \\ \hline
\thead{\makecell{Modular entropy in \\ \textbf{microcanonical} \\ ensembles }} &
   \multicolumn{2}{m{5cm}|}{\centering \vspace{6em} It approximately linear growth $(\sim \log k)$ and transits smoothly at the Page time. \vspace{0.5em}} &
  \makecell{It reaches a maximum saturation \\ value ($S_0=10$) that is independent \\of the replica parameter $n$.} \\ \cline{1-1} \cline{4-4}
\thead{\makecell{Modular entropy in \\ \textbf{canonical} \\ ensembles}} &
  \multicolumn{2}{m{5cm}|}{}&
  \makecell{It also attains a saturation value \\ \big(related to $\frac{1}{n \beta}$ \big) that decreases \\as  $n$ or the inverse temperature $\beta$} \\ \hline
  \thead{\makecell{CoE  in \\ \textbf{microcanonical} \\ ensembles }} &
   \multicolumn{2}{m{5cm}|}{\centering \vspace{6em} It exhibits exponential growth until it reaches its peak at the Page time. The peak value increases as $n$ increases. \vspace{0.5em}} &
  \makecell{It decays to zero.} \\ \cline{1-1} \cline{4-4}
\thead{\makecell{CoE  in \\ \textbf{canonical} \\ensembles }} &
  \multicolumn{2}{m{5cm}|}{}&
  \makecell{It decays to a non-zero value. \\ This value decreases as $n$ or \\ $\beta$ increases.} \\ \hline
\end{tabular}.
\caption{The behavior of modular entropy and CoE at different stages of evaporation for the EoW model.}
\label{table2}
\end{table}

\section{Replica Parameter Corrections to the Island Model} \label{sec3}
\quad Up to now, we have calculated the time evolution of modular entropy and CoE in the EoW model by the replica trick. However, the EoW model is just a simple toy model. In particular, for the canonical ensemble, we have only provided numerical results in \mpref{modular canonical} and \mpref{coe canonical}. Despite providing an analysis of the behavior of the modular entropy and the CoE at early and late times (\eqref{smod canconical early}, \eqref{smod canonical late}, \eqref{cn canonical early}, \eqref{cn canonical late}) using reasonable approximations, this approach remains inadequate. Since this  approximation is not applicable to the behavior of entropy and CoE near the Page time. Therefore, it is imperative that we extend our study to a more realistic black hole background. For the purpose of grasping more details and the analytical expressions of the modular entropy and the CoE, we now use the replica wormholes method in this section to recalculate entropy and CoE in the eternal two-sided JT black hole copuled to a thermal bath. We comprehensively analyze the detailed behavior of modular entropy and CoE through the whole evaporating process and establish the relationship between the replica parameter $n$ and the inverse temperature $\beta$ (\mpref{disks}). Ultimately, based on this relation, we derive the extended island formula \eqref{island formula n} for JT gravity.

\subsection{Generalized Modular Entropy}  \label{generalized entropy mod}
\begin{figure}[htb]
\centering
\includegraphics[scale=0.35]{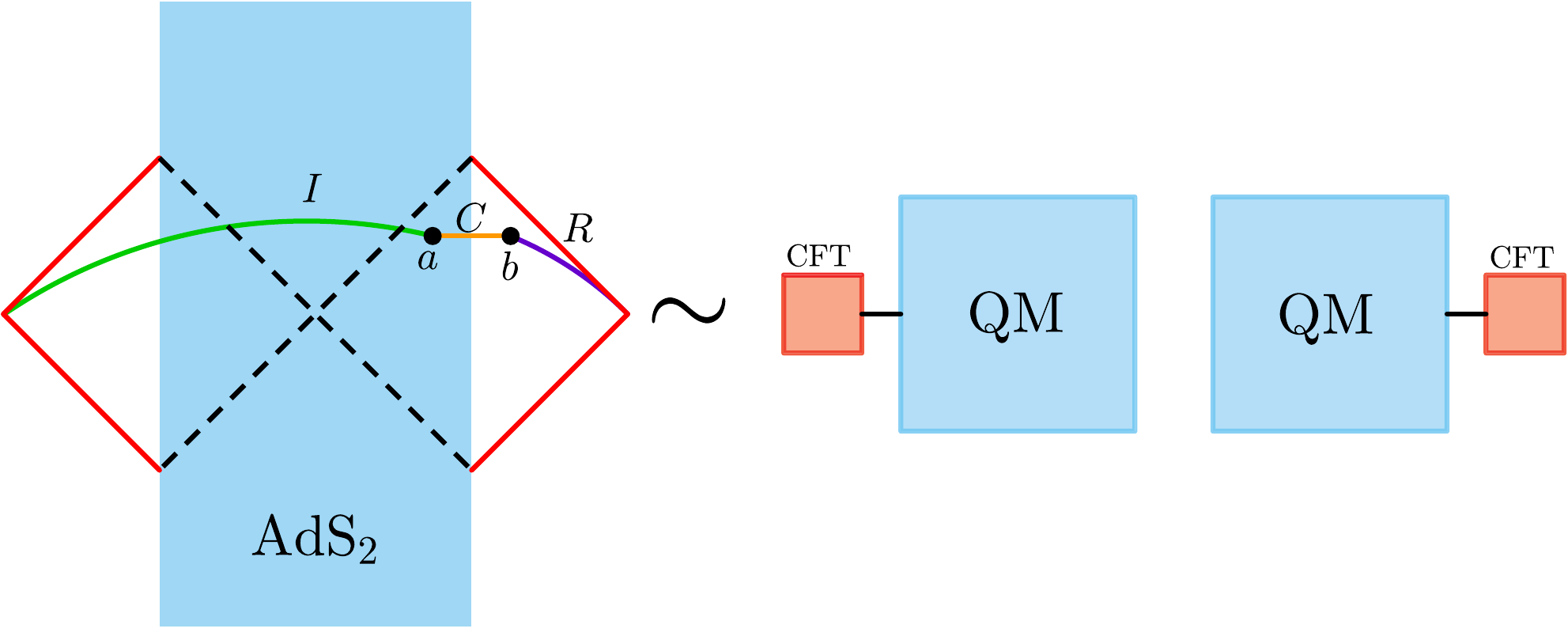}
\caption{\label{jtbath} Left: The JT black hole in thermal equilibrium with the bath. The total system is in a full Hartle-Hawking state. The black hole region is shown in blue; the bath region is shown in red. The two points $a$ and $b$ represent the boundary of the island and the cut-off surface, respectively. The region $I$ and the region $R$ represent the island and radiation, respectively. The region $C$ is their complement. Right: This Hartle-Hawking state is dual to a thermofield double state of the two quantum mechanical systems plus two CFTs. The CFT is located in the Minkowski vacuum.}
\end{figure}

\quad In the beginning, we briefly introduce a concise background and notations. Following the analytically solvable black hole model in \cite{replica2,eternalbh}, we consider the JT gravity in an AdS$_2$ black hole and couple an external flat spacetime with non-gravitational effects. The auxiliary spacetime acts as a thermal bath. Besides, we apply the transparent boundary condition on the boundary between the gravitational region and the non-gravitational auxiliary bath. In this setting, the time evolution of the black hole is observed through the entanglement entropy of the region of radiation $R$, where one can collect the Hawking quanta. See \mpref{jtbath}

\par In the original island proposal \cite{island rule,review}, we inherited the spirit of QES \cite{QES} to calculate the generalized entropy of radiation $S_{\text{gen}}$. Then we obtain the location of QES by extremizing it. Through the competition between the two candidate QES, the minimization requirement limits the entropy of radiation to satisfy the unitary Page curve. This leads to the following island formula in the limit of $n \to 1$
\begin{equation}
\begin{split}
S_R \big |_{n =1} &= \text{min} \Big[ \text{ext} \big( S_{\text{gen}} \big|_ {n=1} \big) \Big] = \text{min} \bigg[ \text{ext} \bigg( \frac{\text{Area}(\partial I)}{4G_N} \bigg|_{n=1} +S_{n=1}^{\text{CFT}} (R \cup I) \bigg) \bigg]  \\
&=\text{min} \bigg[ \text{ext} \bigg( \sum_{\partial I} (S_0 + \phi |_{n=1} (\partial I)) + S_{\text{vN}} (R \cup I) \bigg) \bigg]. \label{island formula2}
\end{split}
\end{equation}
The last line is for the JT gravity version, where $S_0$ is the topological term representing the extremal entropy, and $\phi$ is the dilaton. Now, our goal is to apply the island proposal to the replica geometry. The island appears at late times, which means that the replica geometry turns into $n$ replica wormholes that are cyclically glued along the island region. For this case, one may naively convert the generalized entropy in the island formula \eqref{island formula2} into the following version of \emph{R\'enyi} entropy
\begin{equation}
S_{\text{gen}}(n) = \sum_{\partial I} (S_0 + \phi_n(\partial I)) +S_n^{\text{CFT}} (R \cup I). \label{renyi generalized entropy}
\end{equation}
However, the expression \eqref{renyi generalized entropy} will raise two potential threats \cite{renyi wormholes}: First and foremost, for 2D JT gravity, the dilaton $\phi$ plays the area role of the black hole. Thus the expression \eqref{island formula2} can be regarded as a JT version of the generalized (H)RT formula \cite{RT,HRT} in holography. On the contrary, the replica parameter $n$ extension of the RT formula is \emph{not} actually related to the \emph{R\'enyi entropy}, but to the \emph{modular entropy}. It was first discovered by Xi Dong \cite{modular holo}, who proved that all R\'enyi entropy satisfies a similar area law in holography, as in the RT formula, and are determined by the area of the dual cosmic branes
\begin{equation}
S_{\text{mod}} \equiv n^2 \partial_n  \bigg( \frac{n-1}{n} S_n  \bigg) = \frac{\text{Area}(\text{Cosmic Brane})}{4G_N}. \label{renyi holo}
\end{equation}
Therefore, the first difference is that the area term in the expression \eqref{renyi generalized entropy} dose not match in holography. Secondly, the island is given by the QES condition which is obtained by extremizing the generalized entropy. This calculation involves the variation of the area (dilaton) with respect to the off-shell positions of punctures: $\partial_{a_i} \phi(a_i)$, where $\{a_i\}$ is a \emph{local} coordinate frame. But the second difference is that when we use the QES condition for the expression \eqref{renyi generalized entropy}, it involves the derivative of the area evaluated at the punctures: $\partial_{w_i} \phi(w_i) |_{w_i=a_i}$, where $\{w_i\}$ is a \emph{global} coordinate frame. These two sets of coordinate frames are described in detail later.

\par Based on two points above, we can reasonably conclude that the correct entropy of radiation of the $n$-extension should be taken the following form
\begin{equation}
S_{\text{gen}}(n) = S_{\text{gen}}^{\text{mod}} = \sum_{\partial I} (S_0 + \phi_n(\partial I)) +S_{\text{mod}}^{\text{CFT}}(R \cup I). \label{modular generalized entropy}
\end{equation}
Once we identify the dilaton $\phi$ as the area, the above expression is fully consistent with the holographic R\'enyi entropy \cite{modular holo}. The strict proof of this formula can refer to \cite{renyi wormholes}. Also see appendix \ref{appendixb} for brief proof.

\subsection{Gravitational Path Integral for Modular Entropy and CoE} \label{path integral}
\quad We now provide a more concise derivation of the modular entropy and the CoE by using the gravitational path integral in the island model. We finally obtain the analytical expressions that differ from the numerical result (\mpref{modular canonical} and \mpref{coe canonical}) in the EoW model, which provides a clearer and more comprehensive representation of evolutions for the modular entropy and the CoE. Consider the JT gravity coupled to a CFT with the large central charge: $1 \ll c \ll \frac{1}{G_N}$. The explicit calculation of modular entropy involves a replica manifold $\mathcal{\tilde{M}}_n$, which is completely fixed in the region without gravity. While in the gravitational region, only manifolds with any topology can be considered. They follow certain boundary conditions. In the semi-classical limit, we can decouple the matter part from the dynamic gravitational part and maintain the dominant contribution from the gravity. Then we have the full action that is the sum of the gravitational part and the partition function $Z_n$ of the matter part on the manifold $\mathcal{\tilde{M}}_n$ \cite{replica2}
\begin{equation}
\frac{\log Z_n}{n} = -\frac{1}{n} I_{\text{grav}} [\mathcal{\tilde{M}}_n] + \frac{1}{n} \log \bigg( Z_{\text{CFT}} [\mathcal{\tilde{M}}_n] \bigg). \label{partition function action}
\end{equation}
For JT gravity \cite{JT1,JT2},
\begin{equation}
\begin{split}
-\frac{1}{n} I_{\text{grav}} &= - \frac{1}{n}  \Big( I_{\text{JT}} + I_{\text{EH}} + I_{\text{bdy}} \Big) \\
                             &= \frac{1}{4\pi} \int_{\Sigma} \phi (R+2) + \frac{S_0}{4\pi} \int_{\Sigma} R + \frac{S_0}{2\pi} \int_{\partial \Sigma} \mathcal{K} - \frac{n-1}{n} \mathcal{A}(n). \label{jt action2}
\end{split}
\end{equation}
Here we set $4G_N \equiv 1$ for convenience and set the cosmological constant $\Lambda =2$ to fit the AdS$_2$. $\mathcal{A}(n)$ is denoted as a area term of the conical singularities. Then we use the replica trick to calculate the modular entropy
\begin{equation}
S_{\text{mod}} = (1- n \partial_n) \log \Big[ \text{Tr} (\rho^n) \Big] = \partial_{\frac{1}{n}}  \bigg[ \frac{1}{n} \log \Big[ \text{Tr} (\rho^n) \Big] \bigg]. \label{modular entropy2}
\end{equation}
For the calculation of $\text{Tr} (\rho^n) $, it is actually to evaluating the partition function$Z_n$ on the replica manifold $\tilde{\mathcal{M}}_n$. We assume that the theory has a replica $\mathbb{Z}_n$ symmetry, and consider another manifold $\mathcal{M}_n \equiv \frac{\tilde{\mathcal{M}}_n}{\mathbb{Z}_n}$. This orbifold
$\mathcal{M}_n$ can be regarded as a manifold in which $n$ identical copies of the field theory exist. Besides, one needs to note that $\tilde{\mathcal{M}}_n$ has the conical singularity only in the region without gravity and is smooth in the region with gravity. But $\mathcal{M}_n$ is exactly the opposite. It has conical singularities in the coupled gravitational region and is smooth in the non-gravitational region. The effective action of $\tilde{\mathcal{M}}_n$ and $\mathcal{M}_n$ are related as follows \cite{replica2}
\begin{equation}
\begin{split}
\frac{1}{n} I_{\text{grav}} [\tilde{\mathcal{M}}_n] &= I_{\text{grav}} [\mathcal{M}_n] + \frac{n-1}{n} \mathcal{A}(n) \\
                                                    &= I_{\text{grav}} [\mathcal{M}_n] + \frac{n-1}{n} \sum_i \Big[ S_0 +\phi_n (w_i) \Big]. \label{eff action}
\end{split}
\end{equation}
The derivation of the second step is due to the fact that the area term $\mathcal{A}(n)$ is the sum of the topological constant $S_0$ which comes from the topology of $\mathcal{M}_n$ and the value of the dilaton $\phi$ at the conical singularity. $w_i$ is denoted as the positions of the conical singularity. These positions can be derived from the EOM that obtained the action on the orbifold $\mathcal{M}_n$ with respect to positions
\begin{equation}
- \frac{n-1}{n} \partial_{w_i} \phi(w_i) + \partial_{w_i} \bigg[ \frac{1}{n}  \log \big( Z_{\text{CFT}} (\tilde{\mathcal{M}}_n) \big) \bigg]=0. \label{jt eom1}
\end{equation}
Actually, this expression can be derived to the QES condition for the modular generalized entropy \eqref{modular generalized entropy}
\begin{equation}
\partial_{w_i} \Big[S_0+ \phi_n (w_i) +S_{\text{mod}}^{\text{CFT}} \Big]=0, \label{qes condition}
\end{equation}
where we used the relation: $\frac{1}{n}  \log \big( Z_{\text{CFT}} (\tilde{\mathcal{M}}_n) \big) \equiv S_{\text{mod}}^{\text{CFT}}$.

\par Further, according to the relation \eqref{eff action}, we obtain the partition function $Z_n$ on the manifold $\tilde{\mathcal{M}}_n$ is
\begin{equation}
\begin{split}
-\frac{1}{n} \log \Big[ \text{Tr} (\rho^n) \Big] &= \frac{1}{n} I_{\text{grav}} [\mathcal{\tilde{M}}_n] - \frac{1}{n}  \log \big( Z_{\text{CFT}} (\tilde{\mathcal{M}}_n) \big) \\
                                                 &= I_{\text{grav}} [\mathcal{M}_n] + \frac{n-1}{n} \mathcal{A}(n) - \frac{1}{n}  \log \big( Z_{\text{CFT}} (\tilde{\mathcal{M}}_n) \big). \label{jt eom2}
\end{split}
\end{equation}
By combing it with the expression for modular entropy \eqref{modular entropy2}, we finally obtain
\begin{equation}
\begin{split}
S_{\text{mod}} &= \frac{1}{n} \frac{\delta I^{\text{tot}}}{\delta g_{\mu \nu}} \partial_{\frac{1}{n}} g_{\mu \nu} - \frac{1}{n}  \frac{\delta I_{\text{grav}}[\mathcal{\tilde{M}}_n]}{\delta \phi} \partial_{\frac{1}{n}} \phi + \partial_{\frac{1}{n}} \Big[ \frac{1}{n} \log \big( Z_{\text{CFT}} (\tilde{\mathcal{M}}_n) \big) \Big] \bigg|_{g} + \mathcal{A}(n) \\
               &=\sum_i [S_0 + \phi_n (w_i)] +S_{\text{mod}}^{\text{CFT}}, \label{modular entropy3}
\end{split}
\end{equation}
where $I^{\text{tot}}$ is the total action
\begin{equation}
I^{\text{tot}} \equiv \log (Z[\tilde{\mathcal{M}}_n]) = -I_{\text{grav}} [\tilde{\mathcal{M}}_n] + \log \big( Z_{\text{CFT}} [\tilde{\mathcal{M}}_n] \big), \label{total action}
\end{equation}
and $S_{\text{mod}}^{\text{CFT}}$ is the modular entropy for the matter part
\begin{equation}
S_{\text{mod}}^{\text{CFT}} \equiv \partial_{\frac{1}{n}}  \bigg[ \frac{1}{n} \log (Z_{\text{CFT}} [\tilde{\mathcal{M}}_n]) \bigg].  \label{cft modular entropy}
\end{equation}
The final expression \eqref{modular entropy3} is consistent with the modular generalized entropy \eqref{modular generalized entropy}. For the limit $n \to 1$, the modular entropy is reduced to the von Neumann entropy. Then the expressions \eqref{qes condition}, \eqref{jt eom2}, and \eqref{modular entropy3} are recovered the original expression \eqref{island formula2} in the island proposal \cite{island rule}. Therefore, our derivation is right. Next, the CoE is given by taking derivative of the expression \eqref{modular entropy3} with respect to $n$. To rigorous define the CoE $C_n$ as a derivation of the modular entropy \eqref{modular entropy2}, we assume that the saddle used to calculate $S_{\text{mod}}$ remains smooth under an infinitesimal transformation of $n$. Then
\begin{equation}
\begin{split}
C_n &= -n \partial_n S_{\text{mod}} \\
    &=-n \sum_i \partial_n \phi_n(w) \big|_{w=w_i} +C_n^{\text{CFT}} - n \sum_i \partial_{w_i} \Big[ \phi_n (w_i) +S_{\text{mod}}^{\text{CFT}} \Big] \partial_n w_i, \label{COE2}
\end{split}
\end{equation}
where we used the following relation to recast
\begin{subequations}
\begin{align}
\partial_n \mathcal{A}(n) &= \sum_i \bigg( \partial_n \phi_n (w) \big |_{w=w_i} + \sum_j \frac{\partial \phi_n (w_i)}{\partial w_i} \frac{\partial w_i}{\partial n} \bigg), \label{singularity} \\
         C_n^{\text{CFT}} &= -n \partial_n S_{\text{mod}}^{\text{CFT}}. \label{cft COE}
\end{align}
\end{subequations}

\par To sum up, we obtain the final expression for the modular entropy \eqref{modular entropy3} and the CoE \eqref{COE2}. Before we begin the explicit calculation, the following points need to be emphasized. First, although we can determine that which saddle dominates the black hole evaporation by the entropy \eqref{modular entropy3} based on the generalized modular entropy \eqref{modular generalized entropy}, no similar method exists for the CoE \eqref{COE2}. The expression of CoE \eqref{COE2} is directly applied to the selected saddle of the island formula \eqref{island formula2}. Furthermore, as in the expression \eqref{COE2}, the position of the conical singularity $w_i$ in the expression \eqref{COE2} is also determined by the QES condition \eqref{qes condition}. Most importantly, there is a tricky conformal welding problem \cite{replica2}, which complicates our explicit calculation for \eqref{modular entropy3} and \eqref{COE2}. We discuss it and its resolution in the next subsection. We also give a more elaborate derivation of expressions \eqref{modular entropy3} and \eqref{COE2} in appendix \ref{appendixb}.

\subsection{Conformal Welding Problem and High Temperature Limit}  \label{conformal welding}
\begin{figure}[htb]
\centering
\includegraphics[scale=0.35]{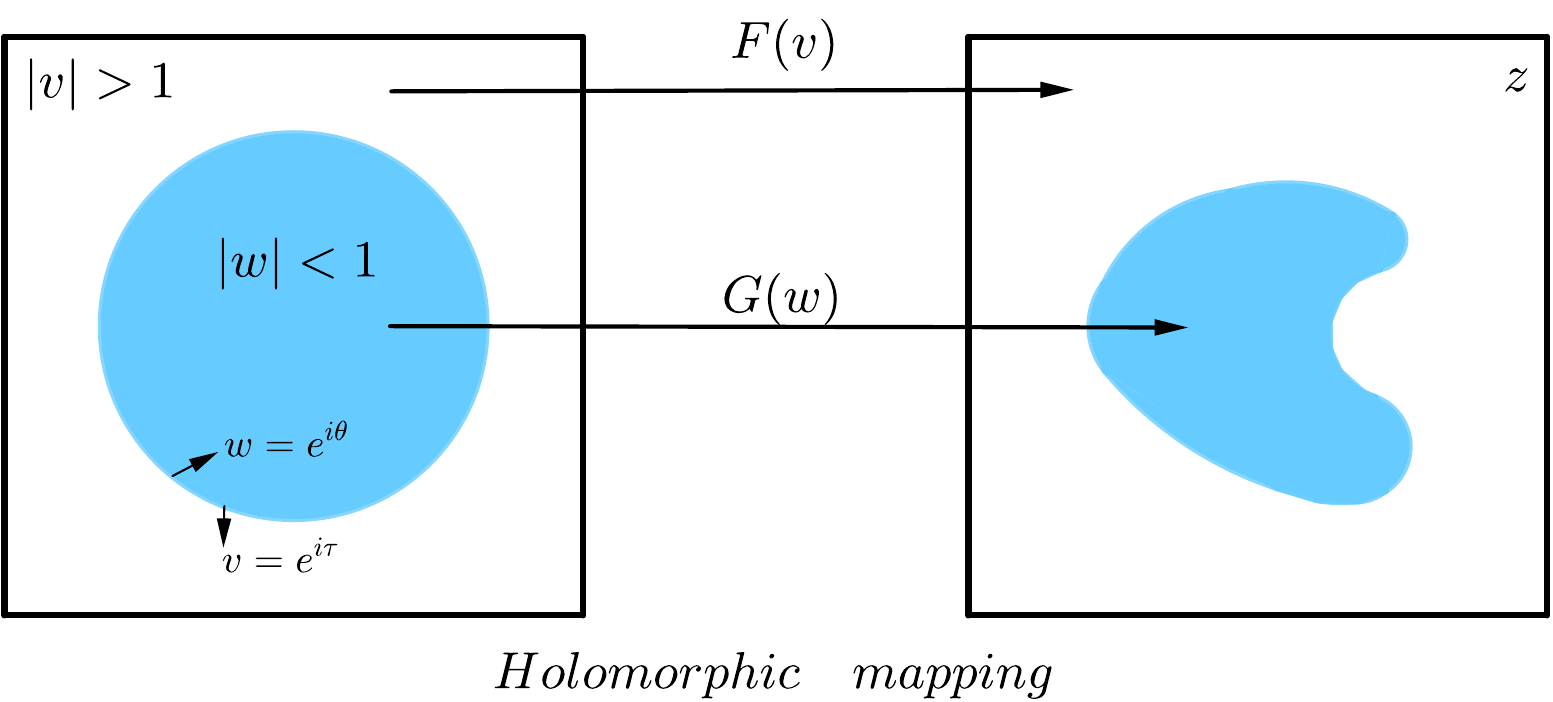}
\caption{\label{mapping} Schematic diagram of the conformal welding problem. There are two disks (shown in blue), parameterized by $|w| \le 1$ and $|v| \ge 1$. Their boundaries are glued by the given function where $w=e^{i \theta}$ and $v=e^{i \tau}$. Two holomorphic functions $G(w)$ and $F(v)$ satisfying conditions \eqref{function G} \eqref{function F} \eqref{function bdy} map each disk onto the complex plane $z$.}
\end{figure}

\quad In this section, we elaborate on the conformal welding problem arising in the explicit calculation and its solution. The conformal welding refers the process of cutting a circle on two Riemann manifolds and gluing them smoothly on their boundaries to obtain a new (Riemann) manifold from the original two manifolds. We use complex analysis to concretize this process. Given two Riemann manifolds, each of which has a disk on it (see \mpref{mapping}). One of them is parameterized by $|w| \le 1$ and the other is parameterized by $|v| \ge 1$. Their boundaries are located as $|w|=1$ and $|v|=1$. However, we can not extend the boundary to the holomorphic map inside the disk. In spite of this, we can still find another coordinate frame ${z}$, which ensures that there is a holomorphic mapping form $|w| \le 1$ and $|v| \ge 1$ to the coordinate ${z}$. Namely, we can probably find two functions $G$ and $F$ that satisfy \cite{replica2}
\begin{subequations}
\begin{align}
z&=G(w), \qquad |w|\le 1 \qquad  \qquad (\text{inside the disk}). \label{function G}\\
z&=F(v), \qquad \  |v|\ge 1 \qquad \qquad (\text{outside the disk}). \label{function F}\\
G(e^{i \theta(\tau)})&= F(e^{i \tau}), \ \ \ \  |w|=|v|=1 \quad \ \ (\text{at the boundaries}). \label{function bdy}
\end{align}
\end{subequations}
The functions $G$ and $F$ are holomorphic on their respective domains but need not satisfy this requirement on the boundary. Finding the functions $F$ and $G$ that satisfy these conditions \eqref{function G}, \eqref{function F}, \eqref{function bdy} is called the ``conformal welding problem''. Functions $F$ and $G$ ultimately depend non-locally on the boundary mode $\theta(\tau)$ and they map the inside of outside of the disk to the inside and outside of some irregular regions on the complex plane $z$ (\mpref{mapping}). One can refer to for more discussion about the conformal welding problem \cite{conformal welding}.

\par In the background of the AdS$_2$ coupled with an auxiliary bath as mentioned before, $w$ represents the AdS$_2$ region with JT gravity and $v$ represents the bath region. The conformal welding problem becomes simple in this background, since the imposed transparent boundary condition can glue the two regions smoothly. However, the computational difficulties that arise when we use expressions \eqref{modular entropy3} and \eqref{COE2} to calculate the entropy and CoE can complicate the conformal welding problem. Concretely, we evaluate $\text{Tr} (\rho^n)$ by the replica trick to obtain the entropy equivalent of the insertion of a twist operator in CFT of the bath region without gravity. In addition, for the AdS$_2$ region containing gravitational effects, the non-perturbative effect dynamically induces an extra conical singularity. In the case of conical singularities, the gluing process becomes non-trivial. These singularities have a back-reaction on the geometry of the gravitational region, which leads to an $n$-dependent EMO at the boundary.

\par
\begin{figure}[htb]
\centering
\includegraphics[scale=0.12]{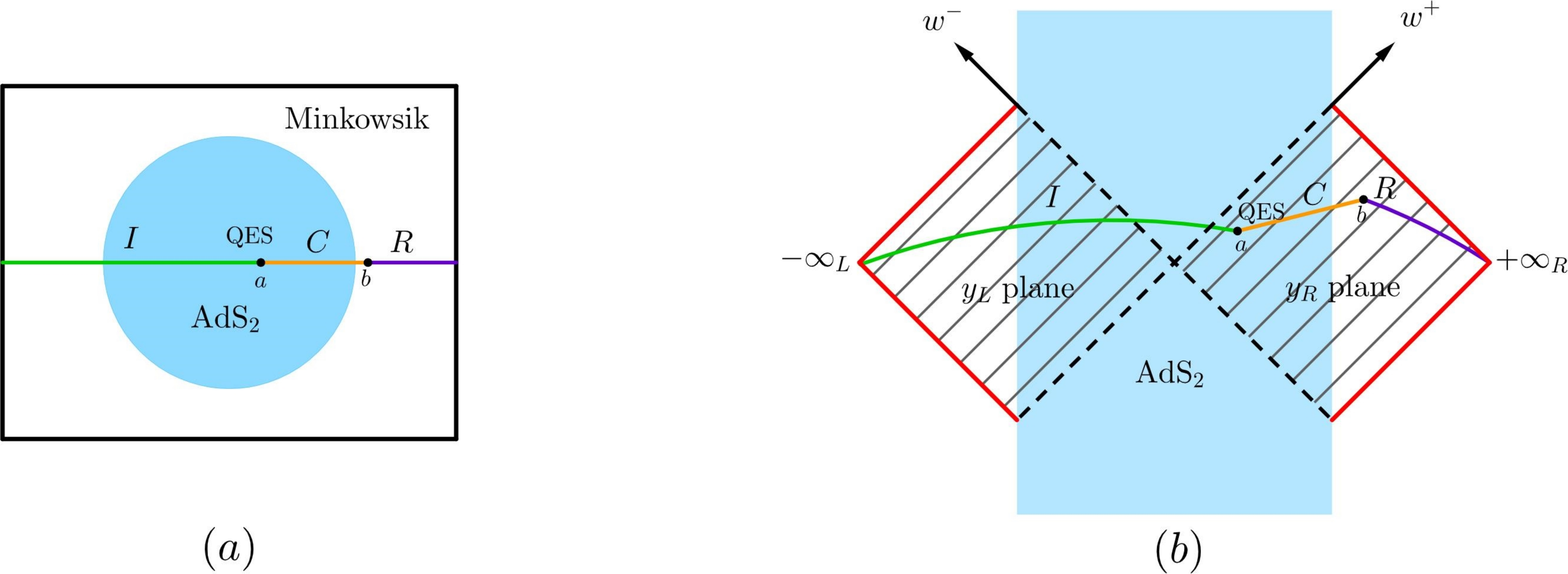}
\caption{\label{signature} The Euclidean and Lorentzian signatures for JT black holes. The island region and the radiation region are labeled $I$ and $R$, respectively. Their endpoints are a QES and the cut-off surface is $b$. The region $C$ is the complementary of $I \cup R$. (a) On the left, the disk represents the JT gravity in AdS$_2$. The flat Minkowski spacetime is coupled to the boundary of the disk. (b) On the right, blue represents the gravitational region, while red is the non-gravitating bath. $\pm \infty_{R/L}$ is denoted as the future/past spacelike infinity and $\{w^{\pm}\}$ is denoted as a global coordinate. The shaded region in the left and right wedges covered by the local coordinates $\{y^{\pm}\}$.}
\end{figure}

 In order to display more concretely the difficulty of the conformal welding problem, we now give the explicit procedure of calculations. Consider the construction in \mpref{signature}. The simplest case is that there is only one QES in the JT black hole (see \mpref{signature}). One the one hand, the calculation for one QES scenario is intuitive and easy to handle. One the other hand, we do not worry about the back-reaction from the other QES to a particular one. Nevertheless, the expected answer can still be obtained in this simplest case\footnote{We will briefly discuss the case of multi QES configurations at the section \ref{sec4}. For more concrete content, see \cite{renyi wormholes}.}.

\par We pay attention to the region $C$, since our goals \eqref{modular entropy3} and \eqref{COE2} relate to the contribution of matter fields on it. The primary step in calculating them is to obtain a conformal welding map with $n$-dependence. This is subject to the modular entropy for CFT \cite{EE formula}
\begin{equation}
S_{\text{mod}}^{\text{CFT}} = \frac{c}{6n} \log \bigg[ \frac{(F(b)-G(a))^2}{\epsilon^2_{\text{UV}} F^{\prime}(b) G^{\prime}(a) \Omega(a) \Omega(b)} \bigg], \label{modular matter part}
\end{equation}
where $\epsilon^2_{\text{UV}}$ is the UV cutoff, $\Omega$ represents the conformal factor at points $a$ and $b$. In the limit $n \to 1$, the conformal map becomes trivial, i.e., $F=G=1$. Then the expression \eqref{modular matter part} is just the simple logarithmic law of the entanglement entropy ($n \to 1$): $S=\frac{c}{6} \log \frac{\ell}{\epsilon_{\text{UV}}}$, where $\ell$ is the length of the system. However, for the general $n$, the modular entropy and the CoE depend on $n$ by terms $F$ and $G$ as well as $\partial_n F$ and $\partial_n G$. Therefore, this requires us to deal with the conformal welding problem and find the final analytic expression of the map \eqref{function G} and \eqref{function F}. Fortunately, although the EOM of the dilaton is complicated for finite $n$, we can still obtain some non-trivial result in the \emph{high temperature limit}: $\beta \sim \kappa \simeq 0$, where $\kappa$ is a dimensionless combination
\begin{equation}
\kappa \equiv \frac{c \beta G_N}{24 \pi \phi_r} \ll1 . \label{coupling}
\end{equation}
In fact, this corresponds to weak gravitational coupling. In the leading order, we set $F=G$ and ignore the conformal welding effect. Then, the holomorphic functions are determined by \cite{replica2}
\begin{equation}
F(v)= \frac{v}{b-v} + {\cal O}(\kappa v^{-1}), \qquad G(w)= \frac{w}{b-w} + {\cal O}(\kappa w).  \label{holomorphic function}
\end{equation}

\subsection{Modular Island in JT Black Holes} \label{modular island}
\quad Based on the support and specific descriptions of the previous subsections, we can now present explicit calculations here.  We first focus on the limit of $n \to 1$, then extend to the case of general $n$. The metric of JT black holes in the covering space is the AdS$_2$, which is written as
\begin{equation}
ds^2 \big|_{n=1} = -\frac{4 dx^+ dx^-}{(x^- - x^+)^2}, \qquad \phi=\frac{2\phi_r}{(x^--x^+)}, \label{jt metric1}
\end{equation}
with the Poincar$\acute{\text{e}}$ coordinate $x^{\pm} = x^0 \pm x^1$. In order to better serve the later calculation, we obtain the coordinate frame $\{y^{\pm}_{L/R} \}$ for the left and right Rindler wedge (see \mpref{signature} (b)) \cite{eternalbh}
\begin{equation}
y_L^{\pm} = t_L \mp \sigma_L, \qquad y_R^{\pm} = t_R \pm \sigma_R. \label{y coordinate}
\end{equation}
Then the metric of the right wedge \eqref{jt metric1} becomes
\begin{equation}
ds^2 \big|_{n=1}=- \frac{4 \pi^2}{\beta^2}  \frac{dy_R^+ dy_R^-}{\sinh^2 \Big(  \frac{\pi (y_R^- -y_R^+)}{\beta}  \Big)}, \qquad \phi = \frac{2 \pi \phi_r}{\beta}  \frac{1}{\tanh^2 \Big( \frac{\pi}{\beta} (y_R^- - y_R^+)  \Big)}, \label{jt metric2}
\end{equation}
through the relation: $x_R^{\pm} = \tanh \frac{\pi y_R^{\pm}}{\beta}$. Similarly, we can switch to the left wedge by replacing $y_R^{\pm}$ with $y_L^{\pm}$. While for the flat bath region which is coupled to the boundaries of both sides of black holes, we use a cut-off $\epsilon = -\sigma_R$ for the right boundary. Then, the flat bath is smoothly glued to the boundary. Therefore, the metric of the right bath is given by
\begin{equation}
ds_{\text{bath}}^2 = \frac{dy_R^+ dy_R^-}{-\epsilon^2}.  \label{bath metric}
\end{equation}
We now consider the other null coordinate frame $\{w^{\pm} \}$, which can cover the full Cauchy slice that consists of black holes \eqref{jt metric2} and flat baths \eqref{bath metric} spacetime \cite{eternalbh}
\begin{subequations}
\begin{align}
\text{Right Wedge}: w_{\text{JT}}^{\pm} &= \pm e^{\pm \frac{2\pi}{\beta} y_R^{\pm}}, \qquad w_{\text{bath}}^{\pm} = \pm e^{\pm \frac{2\pi}{\beta} (t_R \mp \epsilon)}.  \label{right wedge} \\
\text{Left Wedge}: w_{\text{JT}}^{\pm} &= \mp e^{\mp \frac{2\pi}{\beta} y_L^{\pm}}, \qquad w_{\text{bath}}^{\pm} = \mp e^{\mp \frac{2\pi}{\beta} (t_L \pm \epsilon)}. \label{left wedge}
\end{align}
\end{subequations}
This is similar to the Kruskal transformation for the Schwarzschild spacetime. After this coordinate transformation, the metrics \eqref{jt metric2} and flat baths \eqref{bath metric} transforms into the conformal flat form
\begin{equation}
ds^2 \big|_{n=1} = - \Omega^{-2} dw^+ dw^- , \label{jt metric3}
\end{equation}
with the conformal factors
\begin{subequations}
\begin{align}
\Omega_{\text{JT}} &= \frac{1}{2} (1+ w^+ w^-), \label{conformal factor jt} \\
\Omega_{\text{bath}} &= \frac{2\pi \epsilon}{\beta} \sqrt{- w^+ w^-}. \label{conformal factor bath}
\end{align}
\end{subequations}

\par The next step is to obtain the replica geometry. So we need the metric \eqref{jt metric3} attached to the parameter $n$-dependence. It can be acquired via the replica method gluing together $n$ copies of spacetime and along a set of branch cut\footnote{We can also obtain this by inserting the correlator of twist and anti-twist operator at the location of un-copied geometric branch points \cite{renyi wormholes}.}. For JT gravity, we consider this in the replica geometry $\mathcal{\tilde{M}}_n$ or in the $\mathcal{M}_n$ with the twist operator, in which the Riemann manifold $\mathcal{M}_n$ can be regarded as an $n$-fold cover of the base with respect to the $\mathbb{Z}_n$ symmetry (below \eqref{modular entropy2}). At last, we use the uniformization map from
$\mathcal{\tilde{M}}_n$ to $\mathcal{M}_n$, which maps the replica metric on the base $\mathcal{M}_n$. Then we have the metric with $n$-dependence on the base for black hole region
\begin{equation}
ds_n^2 = \frac{4 |d \tilde{w}|^2}{\big( 1-|\tilde{w}|^2 \big)^2}, \qquad \phi_n = \frac{2\pi \phi_r}{\beta} \frac{1+|\tilde{w}|^2}{1-|\tilde{w}|^2}, \label{jt metric4}
\end{equation}
with the uniformization of the coordinate
\begin{equation}
\tilde{w}^n =w. \label{replica coordinate}
\end{equation}
For the bath region, we impose the following boundary condition
\begin{equation}
\phi \big|_{\infty} = \frac{\tilde{\phi}_r}{\tilde{\epsilon}} \bigg|_{\tilde{\epsilon} \to 0},  \qquad \tilde{w} = e^{\frac{2\pi}{\beta} (- \tilde{\epsilon}+ i \tilde{\theta})} \bigg|_{\tilde{\epsilon} \to 0}, \label{bondary condition2}
\end{equation}
where $\tilde{\phi}_r$ and $\tilde{\theta}$ are the renormalized dilaton and the Euclidean time in the $\{ \tilde{w} \}$ coordinate. Here the Euclidean time is periodic: $\theta \sim \theta + \beta$ by the Wick rotation $t \to i \theta$. Similarly, in the $\{ w \}$ coordinate, we also impose the boundary condition
\begin{equation}
\phi \big|_{\infty} = \frac{\phi_r}{\epsilon} \bigg|_{\epsilon \to 0},  \qquad w = e^{\frac{2\pi}{\beta} (- \epsilon+ i \theta)} \bigg|_{\epsilon \to 0}. \label{bondary condition3}
\end{equation}
By comparing these two boundary conditions, the relationship: $\tilde{\phi}_r = \frac{\phi_r}{n}$ and $\tilde{\epsilon} = \frac{\epsilon}{n}$ is obtained. Since the dilaton is a \emph{scalar}. It is subject to a trivial transformation in the map \eqref{replica coordinate}.
\begin{figure}[htb]
\centering
\includegraphics[scale=0.25]{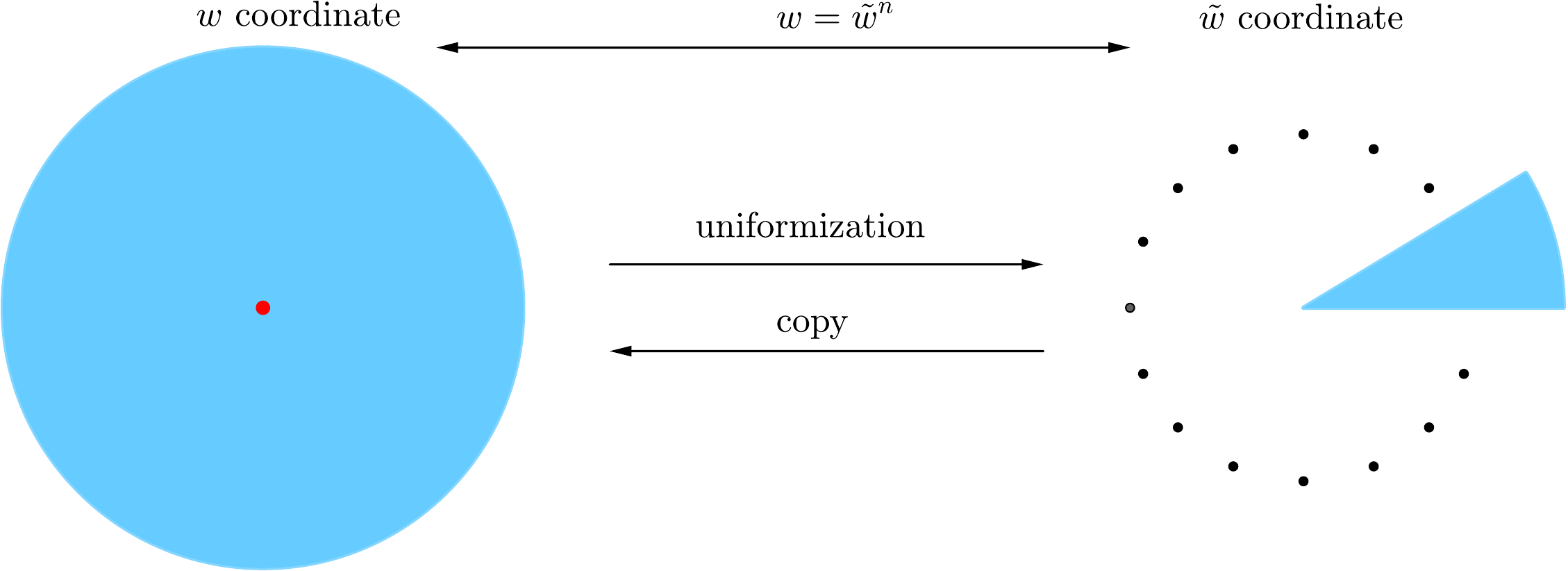}
\caption{\label{singularity1} The two equivalent approaches to describing the conical singularity for the Euclidean signature. On the left, we parameterize the manifold $\mathcal{M}_n$ with the coordinate $w= e^{\frac{2\pi}{\beta} (\sigma + i \theta) }$, where the Euclidean time is periodic $\theta \sim \theta + \beta$ by the Wick rotation $t \to i \theta$, and $\sigma \in (-\infty, - \epsilon)$. On the right, the geometry is uniformized by the coordinate $\tilde{w} = e ^{\frac{2\pi}{\beta} (\tilde{\sigma} + i \tilde{\theta})}$ \eqref{replica coordinate}. The metric is identified with an AdS$_2$ disk with the inverse temperature $\frac{1}{n} \beta$.}
\end{figure}
Additionally, note that the conical singularity is introduced on the $\mathcal{M}_n$ during this process. To determine the dilaton with $n$-dependence in the presence of conical singularities, another equivalent geometric description exists, as depicted in \mpref{singularity1}. Ultimately, the following expression for the dilaton field $\phi_n$ in the context of the conical singularity is derived from \eqref{jt metric4}
\begin{equation}
\phi_n = \frac{2 \pi \tilde{\phi}_r}{\beta} \frac{1+|\tilde{w}|^2}{1- |\tilde{w}|^2} = \frac{2 \pi \phi_r}{n \beta} \frac{1+|w|^{\frac{2}{n}}}{1-|w|^{\frac{2}{n}}} = \frac{2 \pi}{n \beta} \frac{\phi_r}{\tanh \Big( \frac{2 \pi \sigma}{n \beta} \Big)}.  \label{dilaton}
\end{equation}
Therefore, the modular Bekenstein-Hawking entropy is obtained as follows
\begin{equation}
\begin{split}
S_{\text{mod}}^{\text{BH}} &= \frac{\phi_n}{4G_N} \Big|_{\text{horizon}} = \frac{\phi_h}{4G_N}= \frac{\phi_n(w^-=0)}{4G_N} \\
&=S_0 + \frac{\pi \phi_r}{ 2G_N n \beta}, \label{bh entropy}
\end{split}
\end{equation}
where $\phi_h$ is denoted as the value of dilaton at the horizon.

\par We now provide the explicit calculation for the modular entropy and the CoE for the whole system \eqref{jt metric3} through the generalized modular entropy \eqref{modular generalized entropy}. In the case of the non-existence of island\footnote{We show in appendix \ref{appendixc} that modular islands are nonexistent at early times.} $(I = \varnothing)$, only the contribution from radiation in CFT in the single interval $R = [b, +\infty)$ (see \mpref{signature})
\begin{equation}
\begin{split}
S_{\text{gen}}^{\text{no-island}} (n) &= S_{\text{mod}}^{\text{CFT}} (R) \\
                                      &= \frac{c}{6n} \log \Bigg[ \frac{\big( F(b) - F(+ \infty_R) \big)^2}{ \epsilon^2_{\text{UV}}F^{\prime}(+\infty_R) F^{\prime}(b) \Omega_{\text{bath}}(+ \infty_R)  \Omega_{\text{bath}}(b) } \Bigg]. \label{without island1}
\end{split}
\end{equation}
We set the coordinates of points $a$ and $b$ are $a= (t_a,a)$, $b=(t_b,b)$. The coordinate of the spacelike infinity is $+ \infty_R=(t_b, \Lambda \gg b)$, where $\Lambda$ is a IR cutoff. Then the conformal factors \eqref{conformal factor jt} \eqref{conformal factor bath} are read as
\begin{equation}
\Omega_{\text{JT}} (a) = \frac{1}{2} \bigg( 1 - e^{-\frac{4\pi}{\beta} a} \bigg), \qquad \Omega_{\text{bath}} (b) = \frac{2 \pi \epsilon}{\beta} e^{\frac{2\pi b}{\beta}}, \qquad \Omega_{\text{bath}} (+ \infty_R)= \frac{2 \pi \epsilon}{\beta} e^{\frac{2\pi \Lambda}{\beta}}. \label{conformal factors}
\end{equation}
Substituting the holomorphic functions form \eqref{holomorphic function} into the entropy \eqref{without island1}, we obtain the following result at the leading order
\begin{equation}
\begin{split}
S_{\text{mod}}^{\text{no-island}} &= \frac{c}{6n} \log \Bigg[ \frac{d_1^2}{\epsilon^2_{\text{UV}} \ \Omega_{\text{bath}} (+ \infty_R) \ \Omega_{\text{bath}} (b)  } \Bigg] \\
&=\frac{c}{6n} \log \Bigg[ \frac{e^{\frac{4\pi \Lambda}{\beta}} + e^{-\frac{4\pi b}{\beta}} - e^{\frac{2\pi}{\beta} (\Lambda -b)} -e^{\frac{2\pi}{\beta} (\Lambda -b)} }{\epsilon^2_{\text{UV}} \  \frac{4\pi^2 \epsilon^2}{\beta^2} \ e^{\frac{2\pi}{\beta} (b+\Lambda)} }  \Bigg] \\
&\simeq \frac{c \pi }{3n \beta} \log \bigg(  \frac{e^{\Lambda}}{2\pi \epsilon \ \epsilon^2_{\text{UV}}} \bigg) \propto \frac{1}{n \beta} .  \label{smod without island}
\end{split}
\end{equation}
Here $d_1$ represents the geodesic distance between the point $b$ and $+ \infty_R$. In the last line, we retain only the largest term $e^{\frac{4\pi \Lambda}{\beta}}$ in the numerator due to the fact that $\Lambda \gg b$. Further, taking the derivative with respect to $n$, we obtain the CoE as
\begin{equation}
C_n^{\text{no-island}} = -n \partial_n S_{\text{mod}}^{\text{no-island}} =\frac{c \pi}{3n \beta} \log \bigg( \frac{ e^{ \Lambda } }{2\pi \epsilon \ \epsilon^2_{\text{UV}}} \bigg) \propto \frac{1}{n \beta}.  \label{cn without island}
\end{equation}
Since these two expression \eqref{smod without island} and \eqref{cn without island} have both the UV cutoff $\epsilon_{\text{UV}}$ and the IR cutoff $\Lambda$, the results without island is UV and IR divergence. However, since the replica parameter $n$ and the inverse temperature $\beta$ only appear in the denominator, we can naively assume that the larger $n$ or $\beta$, the smaller modular entropy and CoE. It is consistent with our previous numerical results at early times (or before the Page time) (See \mpref{mod3}, \mpref{mod5} and \mpref{coe canonical}).

\par Similarly, we consider the single QES configuration (see \mpref{signature}). The generalized modular entropy with an island is read off as
\begin{equation}
S_{\text{gen}}^{\text{island}} (n) =S_0 + \phi_n(a) +S_{\text{mod}}^{\text{CFT}} (I \cup R). \label{with island1}
\end{equation}
For the matter part at the leading order
\begin{equation}
\begin{split}
S_{\text{mod}}^{\text{CFT}} (I \cup R) &= \frac{c}{6n}  \log \Bigg[ \frac{\big( F(b) -G(a) \big)^2}{\epsilon^2_{\text{UV}} \ F^{\prime}(b) G^{\prime}(a) \Omega_{\text{JT}} (a)  \Omega_{\text{bath}} (b)} \Bigg] \\
&\simeq \frac{c}{6n}  \log \Bigg[ \frac{d_2^2}{\epsilon^2_{\text{UV}} \ \frac{1}{2}  \big(1- e^{- \frac{4\pi a}{\beta}}  \big)  \frac{2\pi \epsilon}{\beta}  \ e^{\frac{2\pi b}{\beta}} } \Bigg] \\
& = \frac{c}{6n}  \log \Bigg[ \frac{\beta \Big( e^{\frac{4\pi b}{\beta}} - e^{ \frac{2\pi}{\beta} (t_b-t_a+b-a)} + e^{-\frac{4\pi a}{\beta}} - e^{\frac{2\pi}{\beta} (t_a-t_b-a+b) }\Big)}{\pi \epsilon \ \epsilon_{\text{UV}}^2 \big( 1-e^{-\frac{4\pi a}{\beta}}  \big) e^{\frac{2\pi b}{\beta}} } \Bigg]\\
&= \frac{c}{6n}  \log \Bigg[ \frac{\beta}{\pi \epsilon \ \epsilon^2_{\text{UV}}}   \frac{\cosh \big[\frac{2\pi}{\beta}(a+b) \big] - \cosh \big[\frac{2\pi}{\beta} (t_a-t_b)\big]}{ \sinh \big( \frac{2\pi a}{\beta} \big)} \Bigg]. \label{with island2}
\end{split}
\end{equation}
Here $d_2$ is labeled as the geodesic distance between the points $a$ (the boundary of the island) and $b$ (the boundary of the cut-off surface). Accordingly, the generalized modular entropy is obtained by \eqref{dilaton} and \eqref{with island2}
\begin{equation}
S_{\text{gen}}^{\text{island}} (n) = S_0 + \frac{2\pi}{n \beta} \frac{\phi_r}{ \tanh \Big( \frac{2\pi a}{n \beta} \Big)} +\frac{c}{6n}  \log \frac{\beta}{\pi \epsilon \ \epsilon_{\text{UV}}^2}   \frac{\cosh \big[\frac{2\pi}{\beta}(a+b) \big] - \cosh \big[\frac{2\pi}{\beta} (t_a-t_b)\big]}{ \sinh \big( \frac{2\pi a}{\beta} \big)}.  \label{with island3}
\end{equation}
At first, we extremize it with respect to time $t_a$, which yield to
\begin{equation}
\frac{\partial S_{\text{gen}}^{\text{island}} (n) }{\partial t_a} = - \frac{c\pi \sinh \big( \frac{2\pi}{\beta} (t_a-t_b) \big)}{3n \beta \Big[ \cosh \big( \frac{2\pi}{\beta} (a+b) \big) - \cosh \big(  \frac{2\pi}{\beta} (t_a-t_b) \big)  \Big]} =0. \label{wrt t}
\end{equation}
The equation indicates that $t_a=t_b$. We substitute this relation into \eqref{with island3} and extremize entropy with respect to $a$ to find the location of the island
\begin{equation}
\frac{\partial S_{\text{gen}}^{\text{island}} (n) }{\partial a} \Bigg|_{t_a=t_b} = - \frac{\pi \bigg[  c n \beta \coth \big( \frac{2\pi }{\beta} a \big) - c n \beta \coth \big( \frac{\pi}{\beta} (a+b) \big) + 12 \pi \phi_r \text{csch}^2 \big( \frac{2\pi}{n \beta} a \big) \bigg]}{3n^2 \beta^2}=0. \label{wrt a}
\end{equation}
The above equation is equivalent to the following condition
\begin{equation}
\frac{12 \pi \phi_r}{ c n \beta} = \frac{1}{2 n \kappa} = \frac{\sinh \big(  \frac{\pi}{\beta} (a-b) \big) \sinh \big( \frac{2\pi}{\beta} a \big) }{\sinh \big(  \frac{\pi}{\beta} (a+b) \big)},  \label{location1}
\end{equation}
where we used the definition \eqref{coupling} to reduce. We take the high temperature limit, in which the gravitational coupling becomes very weak. Namely, $\kappa \ll 1$, and the position of the island is derived from the QES condition \eqref{location1} for the finite $n$\footnote{For the high temperature with the infinite $n$, the QES condition  becomes the indeterminate form. Then the equation \eqref{location1} has no solution at this case and the island can not be found.} is
\begin{equation}
a \to \infty. \label{location2}
\end{equation}
Therefore, the island is located in the center of the AdS$_2$ disk at the high temperature limit, which is exactly at the event horizon. Eventually, the modular entropy with an island is given by substituting this location \eqref{location2} back to \eqref{with island3}
\begin{equation}
\begin{split}
S_{\text{mod}}^{\text{island}} &= S_0 + \frac{2\pi}{n \beta} \frac{\phi_r}{\tanh \big( \frac{2\pi a}{n \beta} \big)} \Bigg|_{a \to \infty} + \frac{c}{6n} \log \Bigg[ \frac{\beta}{\pi \epsilon \ \epsilon^2_{\text{UV}}}  \frac{\cosh \big( \frac{2\pi}{\beta} (a+b) \big)-1}{\sinh \big(\frac{2\pi a}{\beta}  \big)} \Bigg] \Bigg|_{a \to \infty}  \\
&\simeq S_0 + \frac{2\pi \phi_r}{n \beta}  + \frac{c}{6n}  \log \bigg( \frac{\beta}{\pi \epsilon \ \epsilon^2_{\text{UV}}}  e^{- \frac{2\pi b}{\beta}} \bigg) \\
&= S^{\text{BH}}_{\text{mod}} +{\cal O} \bigg(  \frac{b}{n\beta} \bigg) \propto \frac{1}{n \beta}. \label{smod with island}
\end{split}
\end{equation}
Here we used the modular black hole entropy \eqref{bh entropy} and set $4G_N =1$. Next, the CoE with island is
\begin{equation}
\begin{split}
C_n^{\text{island}} &= -n \partial_n S_{\text{mod}}^{\text{island}} = C_n^{\text{thermal}} \\
&= \frac{2\pi \phi_r}{n \beta} + \frac{c}{6n} \log \bigg( \frac{\beta}{\pi \epsilon \ \epsilon^2_{\text{UV}}} e^{-\frac{2\pi b}{\beta}} \bigg) \\
&\simeq \frac{2 \pi \phi_r}{n \beta} +{\cal O} \bigg( \frac{b}{n \beta} \bigg) \propto \frac{1}{n \beta}.  \label{cn with island}
\end{split}
\end{equation}
In the above expressions \eqref{smod with island} and \eqref{cn with island}, the cutoff $\epsilon$ and $\epsilon^2_{\text{UV}}$ are discarded. By comparing them with the results without island \eqref{smod without island} and \eqref{cn without island}, we find that the modular entropy and the CoE with an island are always smaller than the results without island. Correspondingly, the modular entropy and the CoE are limited to a constant for the fixed $n \beta$ . It also implies that the validity of the expressions for the modular entropy \eqref{modular entropy3} and the CoE \eqref{COE2} derived from the gravitational path integral. Further, by comparing the numerical results \eqref{smod canonical late} (\mpref{modular canonical}) and \eqref{cn canonical late} (\mpref{coe canonical}) that we calculated in section \ref{sec2}, we find they are also exactly consistent when set $\phi_r = 2\pi$. Therefore, we also prove the rationality of the numerical result. Crucially, our calculations reveal that both the modular entropy (\eqref{smod without island} and \eqref{smod with island}) and the CoE (\eqref{cn without island} and \eqref{cn with island}) display the coupling $n \beta$, which are approximated to the \emph{thermal entropy} and the \emph{heat capacity}, respectively at late times. This finding not only indicates that the replica parameter $n$ and the inverse temperature $\beta$ exert similar influences on the results, but also implies a deeper connection between modular physical quantities and statistical mechanics (see Table \ref{table1}). We will provide an explanation for this later. In addition, for the limit of $n \to 1$, our results are consistent with \cite{coe1,eternalbh}.
\begin{figure}[htb]
\centering
\includegraphics[scale=0.20]{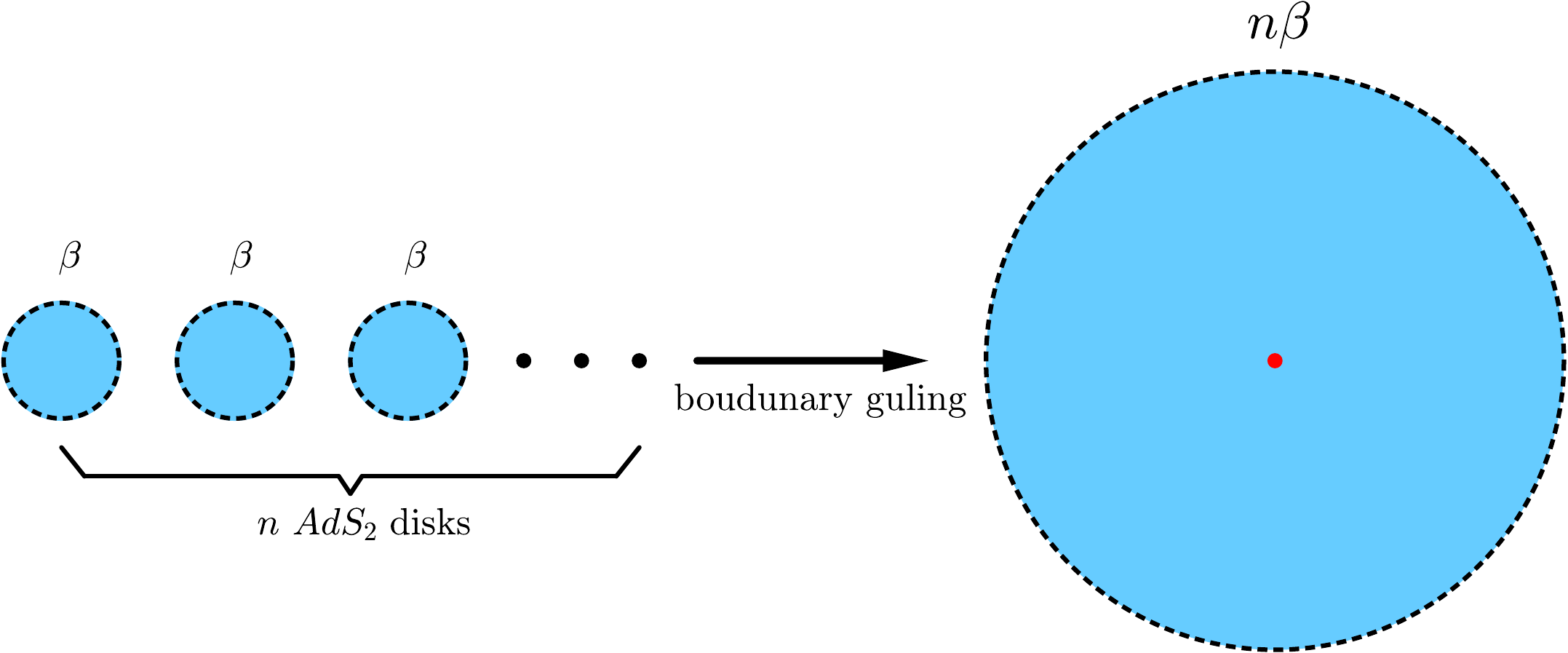}
\caption{\label{disks} At late times, $n$ (copied) AdS$_2$ disks with the inverse temperature $\beta$ are glued along their boundaries to form a \textbf{single} big AdS$_2$ disk with the inverse temperature $n \beta$. In this way, we can describe the properties of these $n$ copied geometries from this \textbf{one disk} alone.}
\end{figure}

\par Now, we compare statistic mechanics to elucidate the  emergence of the coupling $n \beta$ in the final result. As mentioned in the introduction (Table \ref{table1}), in the replica trick, the replica parameter $n$ is analogous to the inverse temperature $\beta$ of the system in statistic mechanics. Based on such analogy, it can be assumed that in the case of the presence of an island or during the replica wormholes stage, the replica geometry ($n$ replicated AdS$_2$ disks) glues along their respective boundaries to form \textbf{one} disk with a inverse temperature $n \beta$ (see \mpref{disks}). If we rescale the inverse temperature by $\tilde{\beta} = n \beta$, then the island phase or the replica wormholes phase is can be described by $\tilde{\beta}$. By bringing this substitution into our results \eqref{smod canonical late}, \eqref{cn canonical late}, \eqref{smod with island} and \eqref{cn with island}, we obtain the result of $\tilde{n}=1$ (a large AdS$_2$ disk composed of $n$ disks glued together). At the same time, this also consistent with the statistical result for $\tilde{n}=1$.  This also indicates that the $n$ replica geometry created by gravity plays a crucial role in the later stage of black hole evaporation. Finally, under this explanation, the island formula \eqref{island formula2} can be matched to the case of general finite $n$ for JT gravity \eqref{island formula n}. Therefore, the relationship between the replica parameter $n$ and the inverse temperature $\beta$ is not merely a simple analogy, but rather an accurate correspondence derived from statistical mechanics.

\section{Discussion and Conclusion} \label{sec4}
\quad In summary, the contribution of this paper is primarily reflected in two aspects. First of all, we have achieved further extension and development of the previous studies \cite{replica1,replica2,eternalbh}. We investigate the situation of general replica parameter $n$ for the replica wormholes in the EoW model and the island model, respectively. We first compute the time-dependent evolution curves of the modular entropy and the CoE under two distinct ensembles in the EoW model. The results are presented in \mpref{modular microcanonical}, \mpref{coe microcanonical}, \mpref{modular canonical} and \mpref{coe canonical}. In particular, for canonical ensembles, the evolution of modular entropy and CoE can only be obtained  through numerical calculations (\mpref{modular canonical} and \mpref{coe canonical}). To verify the accuracy of the numerical calculation, we then derived the approximate analytical expressions \eqref{smod canconical early}, \eqref{cn canonical early}, \eqref{smod canonical late} and \eqref{cn canonical late} at early and late times by using reasonable approximations, which were in agreement with the numerical results. Additionally, we find that the influence of the replica parameter $n$ on the modular entropy and the CoE exhibits a similar pattern to the inverse temperature $\beta$. This motivates us to conduct a more comprehensive investigation into the evolution of black holes in the canonical ensemble, going beyond mere numerical results. Therefore, we turn to the island model to obtain a more accurate and realistic perspective on black hole evaporation. We consider the two-sided eternal JT black hole in thermal equilibrium with the auxiliary bath, which is similar to the case of the EoW model in the canonical ensemble. Based on the holographic duality, we obtain the analytical expressions \eqref{modular entropy3} and \eqref{COE2} for the modular entropy and the CoE through the gravitational path integral. Eventually, we apply these expressions to the JT black hole with a single QES configuration and obtain the explicit results \eqref{smod with island} and \eqref{cn with island}, which are essentially in accordance with the numerical results \eqref{smod canonical late} and \eqref{cn canonical late} at late times.

\par A naturally relevant discussion involves the replica parameter $n$ and its impact. For the EoW model, we summarize the impact of $n$ on the modular entropy and the CoE in Table \ref{table2}. For the island model, it is vital to note that the analytical expressions for both the modular entropy (\eqref{smod without island}, \eqref{smod with island}) and the CoE (\eqref{cn without island}, \eqref{cn with island}) during the evaporation process contain the coupling term $\frac{1}{n \beta}$. This also elucidates the reason for the similar variations in modular entropy (\mpref{mod3}, \mpref{mod5}) and CoE (\mpref{coe canonical}) with respect to $n$ and $\beta$. Furthermore, we can establish a significant connection between statistical mechanics and replica trick method. If we assume that the analogy presented in Table \ref{table1} is feasible, then the replica parameter $n$ should be considered as the inverse temperature $\beta$ of the system. Based on it, a reasonable \emph{physical} interpretation of \emph{mathematical} replica trick can be provided, i.e., at late times, $n$ replica geometries with the inverse temperature $\beta$ ($n$ AdS$_2$ disks) are glued together along boundaries to form a geometry with the inverse temperature $n \beta$ (\mpref{disks}). Then we expect that the replica wormholes phase or the island phase can be described by black holes with an inverse temperature $n \beta$. Under this supposition the modular entropy and the CoE are analogous to the \emph{thermal (Bekenstein-Hawking) entropy} \eqref{smod with island} and the \emph{heat capacity} \eqref{cn with island} of a JT black hole with an inverse temperature $\tilde{\beta} = n \beta$. In the end, we generalize the island formula of JT gravity to the case of finite $n$ by the expression \eqref{island formula n} through this interpretation. In the limit of $n \to 1$, the formula \eqref{island formula n} is reduced the original island formula \eqref{island formula1}. In addition, all of our results reduce to the previous work \cite{eternalbh,replica1,replica2,coe1} when the limit of $n \to 1$ is taken. In short, our calculations provide valuable insights. It offers a set of relevant computations regarding the modular entropy and the CoE and can be applied to some other scenarios.

\par Some other remarkable improvement or deficiencies are as follows. First of all,  we only consider two ensembles in the EoW model, namely, the microcanonical and the canonical ensemble. The grand canonical ensembles with chemical potential, present intriguing possibilities and merit further investigation. In addition, we focus exclusively on the two geometric (the Hawking saddle and the replica wormholes saddle) contributions by the planar approximation. Some relevant studies have demonstrated that the multi-boundary geometry, defined by the junction condition, can adjust the magnitude of the stress-energy tensor at the interface, which potentially rendering this configuration dominant in some cases \cite{junction1,junction2}. Therefore, for the EoW model, this method can also be employed to evaluate the contributions of these saddle points. Secondly, for the conformal welding problem, we only obtain the holomorphic functions $F$ and $G$ by \eqref{holomorphic function} at the leading order through the high temperature limit \eqref{coupling}. We also can consider the contribution from high order terms, which is expected to render the results more precise. Moreover, we only consider the single QES in our calculation for the island model. Although this simplest configuration suffices to yield the unitary result (\eqref{smod with island} and \eqref{cn with island}), the calculation in the island model incorporates the \emph{leading}order. Therefore, the structure with multiple QES should also be considered in detail \cite{renyi wormholes}, as this will enhance the accuracy of the Page curve. Finally and most importantly, it is crucial to emphasize that our study of the relationship between the replica parameter $n$ and the inverse temperature $\beta$ reveals only an analogical association. As illustrated in Table \ref{table1}, a more profound correspondence between statistical mechanics and quantum information theory is suggested, which will constitute the primary focus of our future research endeavors. We leave these points for future work.

\begin{acknowledgments}
We would like to thank Zhen-Bin Yang and Cheng Peng for the helpful discussions on the EoW model and  quantum many-body theories. We also would like to thank Yang Zhou for the valuable discussions on the R\'enyi entropy. The study was partially supported by NSFC, China (Grant No. 12275166 and No. 12311540141).
\end{acknowledgments}

\appendix
\section{Microcanonical Ensemble at the Limit Case}\label{appendixa}
\quad In this appendix, we provide the explicit calculation for the case of $n \to 1$. For the microcanonical ensemble, the result involves the integral expression \eqref{integral2}. It can be expanded by $(n-1)$ as follows
\begin{equation}
\begin{split}
\tilde{I}_2 &= t^{n-1} \cdot \tilde{I}_1  \\
& \simeq t^{n-1} \cdot \bigg[ 1+(n-1) \frac{t}{2} - \frac{(n-1)^2}{2}  \bigg( 1+ \frac{3}{2}t +\frac{1-t^2}{t} \log (1-t) -2 \text{Li}_2(t) \bigg) +{\cal O}(n-1)^3 \bigg], \label{A integral}
\end{split}
\end{equation}
where $\text{Li}_2(t)$ is the Polylogarithm function. Therefore, the von Neumann entropy and the CoE are given by taking the limit $n \to 1$
\begin{equation}
S_{\text{vN}}=S_n |_{n=1} = S_{\text{mod}}|_{n=1} = \left \{
              \begin{array}{lr} S_0 + \log t - \frac{t}{2} , \qquad  \qquad 0 \le t \le 1. &\\
              S_0 - \frac{1}{2t}, \qquad \qquad \qquad \qquad \ \ t \ge 1. &
              \end{array} \right. \label{A vn entropy n1}
\end{equation}
and
\begin{equation}
C_1  = \left \{
              \begin{array}{lr}  \frac{1}{2} t , \qquad  \qquad 0 \le t \le 1. &\\
               \frac{1}{2t}, \qquad \qquad  \qquad t \ge 1. &
              \end{array} \right. \label{A coe n1}
\end{equation}
Finally, we obtain the corresponding curves for $n=1$ in \mpref{modular microcanonical} and \mpref{coe microcanonical} (black lines).

\section{Derivation of Modular Entropy and CoE}\label{appendixb}
\quad In this appendix, we offer an alternative calculation approach to verify our result \eqref{modular entropy3} and \eqref{COE2}. We focus on the Euclidean black hole and set the units of $\frac{\beta}{2 \pi}$ for simplicity. The null coordinate $w$ after the Wick rotation is $w = e^{- r+i\theta}$, which is described inside the AdS$_2$ disk. While the boundary of AdS$_2$ disk is described by the angular coordinate $\theta$ and the time coordinate $\tau$. Therefore, the JT gravity in the $n$ replica geometries can be parameterized by the boundary mode $\theta (\tau)$. Since the replica geometry has $\mathbb{Z}_n$ symmetry. After orbifolding by $Z_n$, there is an additional factor of $n$ and extra terms that produce conical singularities. The effective action becomes \cite{replica2}
\begin{equation}
    \begin{split}
         -\frac{1}{n} I_{\text{grav}}\big [ \tilde{M_n}  \big ] = &\frac{S_0}{4\pi} \left [ \int_{ {\Sigma_{2}} }R + \int_{ {\partial  \Sigma_{2}} }2 \mathcal{K}  \right ] + \int_{ {  \Sigma_{2}} } \frac{\phi}{4\pi} (R+2)+\frac{\phi_b}{4\pi}\int_{ {\partial \sum_{2}} }2 \mathcal{K} \\
         &-\frac{n-1}{n} \sum_{i}\left [ S_0+\phi (w_i) \right ] ,
    \end{split}
    \end{equation}
    \label{B effection action}
where $w_i$ is the position of the conical singularity. Then the action for the boundary mode $\theta (\tau)$ is obtained by substituting the external curvature
\begin{equation}
    -I_{\text{grav}}(n)=-I_{\text{grav}} \big [ \tilde{M_n}  \big ] =S_0+\frac{n\phi_r}{2\pi}\int_{0}^{2\pi} d\tau \left [ \left \{ e^{i\theta},\tau \right \} +\frac{n^2-1}{2n^2}R(\theta)  \right ] ,
    \label{B action 1}
\end{equation}
where $\left \{ x,y \right \} $ is the Schwarzian derivative and
\begin{equation}
    R(\theta) =- \frac{(1-A^2)^2(\partial _{\tau}\theta)^2}{ \left | 1-Ae^{i \theta} \right |^4 }.
    \label{B R}
\end{equation}
The EOM is derived from the variation of the action \eqref{B action 1} with respect to $\tau$
\begin{equation}
    \begin{split}
        -\frac{1}{n}\delta I_{\mathrm{grav}} &= \frac{\phi _r}{2\pi}\int d\tau \left [ \delta \left \{ e^{i\theta},\tau \right \} +\frac{n^2-1}{2n^2} \delta R( \theta )\right ] \\
        & = \frac{\phi _r}{2\pi} \int d\tau \left [ \partial _{\tau}  \left \{ e^{i\theta},\tau \right \}+ \frac{n^2-1}{2n^2} \partial _{\tau} R(\theta)  \right ]  \frac{\delta \theta}{\partial _\tau \theta}.  \label{B variation}
    \end{split}
\end{equation}
Using the Schwarzian composition identity \cite{replica2}
\begin{equation}
    \frac{\phi_r}{2\pi}\left [ \partial _{\tau} \left \{ e^{i\theta},\tau \right \} +\frac{n^2-1}{2} \partial _{\tau} R (\theta) \right ] =i\left ( T_{yy}(i\tau) -T_{\bar{y}\bar{y} } (-i\tau )\right ) ,
    \label{B Schwarzian}
\end{equation}
where $T_{yy}$ and $T_{\bar{y}\bar{y} }$ are stress tensor in the bath region with the coordinate $y$. After solving the conformal welding problem, the holomorphic map $F(v=e^y)$ for bath region is obtained, the EOM is \cite{replica2}
\begin{equation}
    \frac{24\pi \phi _r}{c\beta} \partial _r\left [ \left \{ e^{i\theta},\tau \right \}+\frac{n^2-1}{2 n^2} R(\theta (\tau)) \right ] = ie^{2i\tau}\left [ -\frac{n^2-1}{2n^2} \frac{F^{\prime}(e^{i\tau})^2}{F(e^{i\tau})^2}-\left \{ F,e^{i\tau} \right \}  \right ] +\text{c.c}.
    \label{B eom 1}
\end{equation}
In the high temperature limit $\kappa=\frac{c\beta}{24\pi \phi_n}\sim \beta \sim 0$, the EOM \eqref{B eom 1} reduced to:
\begin{equation}
    \partial _{\tau} \left [ \left \{ e^{i\theta},\tau \right \}+\frac{n^2-1}{2 n^2} R(\theta (\tau)) \right ]=0.
    \label{B eom 2}
\end{equation}
In fact, the ADM mass $M$ is related this equation
\begin{equation}
    M\equiv -\frac{\phi _r}{2 \pi} \left [ \left \{ e^{i\theta},\tau \right \}+\frac{n^2-1}{2 n^2} R(\theta (\tau)) \right ].
    \label{B mass}
\end{equation}
Thus the EOM \eqref{B eom 2} implies the ADS mass is conserved. Then we have to calculate this boundary equation to the leading order in $(n-1)$. We use the definition in \cite{modular entropy} to calculate the modular entropy and the CoE
\begin{equation}
    S_\mathrm{{mod}}=n^2 \partial_n \tilde{I_n}, \qquad C_n = -n\partial _n \left ( n^2 \partial _n \tilde{I_n} \right )  ,
    \label{B expressions}
\end{equation}
with the action
\begin{equation}
    \tilde{I}_n=\frac{1}{n} I_{\mathrm{grav}}(n)-I_{\mathrm{grav}} \big |_{n=1}-\frac{1}{n}\log Z_{\text{CFT}}\big [  \tilde{M_n} \big ] + \log Z_{\text{CFT}}\big [  \tilde{M_1} \big ].  \label{B reduced action}
\end{equation}
Combing the expression \eqref{B action 1} we find
\begin{subequations}
\begin{align}
    \tilde{I}_{\text{grav}} (n)&= \left ( \frac{n-1}{n} \right ) S_0 +\frac{\phi _r}{2\pi} \int_{0}^{2\pi} d\tau \left [ \left \{ e^{i\theta (\tau)},\tau \right \}\bigg|_{n=1}- \left \{ e^{i\theta(\tau)},\tau \right \} \bigg | _n \ \right ]  \notag \\
    & -\frac{\phi _r}{4 \pi} \frac{n^2-1}{n^2} \int_{0}^{2 \pi}  R(\theta)d \tau,
    \label{B grav part} \\
    \tilde{I}_{\text{CFT}} (n)&= -\frac{1}{n} \log Z_{\text{CFT}} \big [  \tilde{M}_n \big ] + \log Z_{\text{CFT}} \big [  \tilde{M}_1 \big ]. \label{B cft part}
\end{align}
\end{subequations}
Then: $\tilde{I}_n=\tilde{I}_{\text{grav}} (n)+\tilde{I}_{\text{CFT}} (n)$. Next, we take the derivative of $\tilde{I}_n$ with respect to $n$
\begin{equation}
    \begin{split}
        \partial _n \tilde{I}_n & = \partial_n \tilde{I}_{\mathrm{grav}} (n)+ \partial_n \tilde{I}_{\mathrm{CFT}} (n)\\
        &= \frac{1}{n^2} S_0 - \frac{\phi _n}{2 \pi} \int_{0}^{2\pi} d\tau \partial _n \left \{ e^{i\theta},\tau \right \} \bigg |_n -\frac{\phi _r}{2\pi n^2}\int_{0}^{2\pi} R(\theta)d \tau -\frac{\phi _r}{4 \pi} \frac{n^2-1}{n^2} \int_{0}^{2\pi} \left [ \partial_n R(\theta) \right ]d\tau \\
        & + \frac{1}{n^2} \log Z_{\text{CFT}}\big [  \tilde{M_n} \big ]-\frac{1}{n} \partial _n \log Z_{\text{CFT}}\big [  \tilde{M_n} \big ] \bigg |_g.     \label{B derivation}
    \end{split}
\end{equation}
Finally, we derive the modular entropy as
\begin{equation}
    \begin{split}
        S_{\mathrm{mod}}&=n^2 \partial _n \tilde{I}_n=S_0-\frac{\phi _r}{2\pi n}\int_{0}^{2\pi} R(\theta)d \tau+\log Z_{\text{CFT}}\big [  \tilde{M_n} \big ]-n\partial_n\log Z_{\text{CFT}}\big [  \tilde{M_n} \big ] \bigg |_g\\
        &-n^2 \frac{\phi _r}{2 \pi}\int_{0}^{\pi} \left [ \partial_n \left \{ e^{i\theta},\tau \right \}-\frac{\phi_r}{4\pi}(n^2-1) \right ] d \tau -\frac{\phi _r}{4\pi}(n^2-1)\int_{0}^{2\pi}\left [ \partial_n R(\theta) \right ]d\tau\\
        & =S_0-\frac{\phi _r}{2\pi n} \int_{0}^{2\pi}R(\theta)d\tau +S_{\text{mod}}^{\text{CFT}}.
    \end{split}
    \label{B result 1}
\end{equation}
In which, we use the EOM \eqref{B eom 2} in the high temperature limit to simplify. Then the CoE is
\begin{equation}
    C_n=-n\partial_n S_{\text{mod}}=-\frac{\phi_n}{2\pi}\int_{0}^{2\pi}R(\theta)\mid _{n=1}d \tau +\frac{\phi_r}{2\pi}\int_{0}^{2\pi}\left [ \partial_n R(\theta)\Big |_{n=1} \right ] d \tau +C_{n}^{\text{CFT}},
    \label{B result 2}
\end{equation}
with $C_{n}^{\text{CFT}}=-n \partial _n S_{\text{mod}}^{\text{CFT}} $.

\par The subsequent goal is to solve the EOM of the boundary mode $R(\theta)$. This process is derived in detail in \cite{replica2}. Here we provide only the necessary derivations. For general $n$ case, the problem has the SL(2,R) gauge symmetry. We can use this gauge to fix $A$ in \eqref{B R}. Then the EOM \eqref{B eom 2} is recast to \cite{replica2}
\begin{equation}
    \partial _\tau \left [  \left \{ e ^{\frac{i\theta \tau}{n}},\tau \right \} +\frac{n^2-1}{2n^2}R(\theta) \right ] =ik e^{2i \tau} \left [ -\frac{n^2-1}{2n^2} \frac{F^{\prime}(e^{i\tau})^2}{F(e^{i\tau})^2} -\left \{F,e^{i\tau} \right \}\right ] +\text{c.c}. \simeq  0.
    \label{B emo 3}
\end{equation}
Since we are need the solution in the order of $(n-1)$, we expand
\begin{equation}
    \theta (\tau)=\tau +(n-1)\delta\theta (\tau).
    \label{B expand}
\end{equation}
Substituting this perturbation into \eqref{B emo 3}, we have
\begin{equation}
    \partial_{\tau} \left( \delta\theta^{\prime \prime \prime} +\frac{1}{n^2} \delta \theta^{\prime} \right)=\frac{4A(1-A^2)^2\sin\tau}{\left |  1-Ae^{i\tau}\right |^6 }.
    \label{B perturbation}
\end{equation}
Then we can expand this in a Fourier series
\begin{equation}
    \delta \theta =\sum_{m=-\infty}^{+\infty } C_m e^{im\tau},
    \label{B forier}
\end{equation}
with the condition for boundary mode $\theta(\tau)$
\begin{equation}
    \theta(\tau+2\pi)=\theta(\tau)+2\pi,\quad-\theta(\tau)=\theta(-\tau),\quad \delta \theta(\tau+2\pi)=\delta \theta(\tau),\quad -\delta\theta(\tau)=\delta\theta(-\tau).
    \label{B condition}
\end{equation}
Thus, the EOM becomes
\begin{equation}
    \sum_{m=-\infty}^{+\infty }C_mm^2(1-m^2)e^{im\tau}=\sum_{m=-\infty}^{+\infty }\frac{2mA^m\left [ 1+A^2(1-m)+m \right ] }{1-A^2}e^{im\tau}.
    \label{B emo4}
\end{equation}
When $m=1$, the left hand of the above equation is zero, while the right hand left a Fourier cofficient: $\frac{2A}{1-A^2}$. So we impose the following condition
\begin{equation}
    \frac{2A}{1-A^2}e^{i\tau}=0.
    \label{B singulartity}
\end{equation}
It turns out that the position of conical singularity coming from the EOM of the boundary mode satisfies $A=0$, which is the origin of AdS$_2$ disk. In this case, the EOM has a simple form
\begin{equation}
    \sum_{m=-\infty}^{+\infty }C_mm^2(1-m^2)e^{im\tau}=0.
    \label{B equation}
\end{equation}
We obtain the solution is (see \eqref{B forier})
\begin{equation}
    \delta \theta (\tau)=c_1 e^{i\tau},
    \label{B solution}
\end{equation}
where $c_1$ is a real number. In the case of $A=0$, the related term in the expansions \eqref{B result 1} and \eqref{B result 2} as follows:
\begin{equation}
    -\frac{\phi_r}{2\pi} \int_{0}^{2\pi} R(\tau)d\tau=\phi_r, \qquad \frac{\phi_r}{2\pi}\int_{0}^{2\pi} \left [ \partial_n R(\theta) \right ]d \tau =-\frac{\phi_r}{\pi} \int_{0}^{2\pi} \partial_{\tau} \left [ \partial_n \theta (\tau) \right ] d\tau=0.
    \label{B terms}
\end{equation}
In the end, the modular entropy \eqref{B result 1} and the CoE \eqref{B result 2} is given by
\begin{equation}
    S_{\mathrm{mod}}=S_0 + \frac{2 \pi \phi _r}{n\beta}+S_{\mathrm{mod}}^{\mathrm{CFT}}.
    \label{B result 3}
\end{equation}
and
\begin{equation}
    C_n =\frac{2 \pi \phi_r}{n\beta}+C_n^{\mathrm{CFT}}.
    \label{B result 4}
\end{equation}
Here we recover the units $\frac{2\pi}{\beta}$. Therefore, these final expressions are consistent with our previous results \eqref{smod with island} and \eqref{cn with island}.

\section{No Modular Islands at the Early Stage}\label{appendixc}
\quad In this appendix, we provide more explicit calculation for modular entropy in JT black holes at the early stage. We show that modular islands are absents at early times, which render our results \eqref{smod without island} and \eqref{cn without island} more persuasive. We assume that island has already existed at early times, thus the generalized entropy is from \eqref{with island3}
\begin{equation}
S_{\text{gen}}^{\text{island}} (\text{early}) = S_0 + \frac{2\pi}{n \beta} \frac{\phi_r}{ \tanh \Big( \frac{2\pi a}{n \beta} \Big)} +\frac{c}{6n}  \log \frac{\beta}{\pi \epsilon \ \epsilon_{\text{UV}}^2}   \frac{\cosh \big[\frac{2\pi}{\beta}(a+b) \big] - \cosh \big[\frac{2\pi}{\beta} (t_a-t_b)\big]}{ \sinh \big( \frac{2\pi a}{\beta} \big)}.  \label{C late entropy}
\end{equation}
Using the approximation at the early stage: $t_a,t_b \ll a,b$. Then the above expression is reduced to
\begin{equation}
S_{\text{gen}}^{\text{island}} (\text{early}) \simeq S_0 + \frac{2\pi}{n \beta} \frac{\phi_r}{ \tanh \Big( \frac{2\pi a}{n \beta} \Big)} +\frac{c}{6n}  \log \frac{\beta}{\pi \epsilon \ \epsilon_{\text{UV}}^2} \frac{e^{\frac{2\pi}{\beta} b} - e^{\frac{2\pi}{\beta} t_a} - e^{\frac{2\pi}{\beta} t_b} }{e^{\frac{2\pi}{\beta}a} -1}.  \label{C approximation}
\end{equation}
The location of islands is given by extremizing the above equation
\begin{equation}
\frac{\partial S_{\text{gen}}^{\text{mod}} (\text{early}) }{\partial t_a} = \frac{c \ \pi \ e^{\frac{2\pi}{\beta} t_a } }{3 \bigg( e^{\frac{2\pi}{\beta}t_a} + e^{\frac{2\pi}{\beta} t_b } - e^{\frac{2\pi}{\beta} b}       \bigg)} =0. \label{C ext t}
\end{equation}
and
\begin{equation}
\frac{\partial S_{\text{gen}}^{\text{mod}} (\text{early}) }{\partial a} = \frac{-\pi \big[ c \ n \ \beta \big(1+\coth(\frac{\pi}{\beta}a )\big) + 24 \pi \ \phi \ \text{csch}^2 \big( \frac{2\pi}{\beta} a \big)   \big]}{6n^2 \ \beta^2}=0. \label{C ext a}
\end{equation}
The equation \eqref{C ext a} has no solution based on the positivity of hyperbolic cosecant function: $\coth x +1 >0$. Therefore, the island is absent at early times, which result in the contribution of modular entropy is exclusively from CFT \eqref{without island1}. This also proves the validity of our result \eqref{smod without island}.

\end{document}